\documentclass[aps,a4paper,prb,amsmath,twocolumn,amssymb,titlepage]{revtex4-1}

\usepackage{graphicx}
\usepackage{amsfonts}

\usepackage{amssymb}
\usepackage{amsmath} 

\usepackage{multirow} 
\usepackage{tabularx} 
\usepackage{array}

\usepackage{braket}

\usepackage{bm}
\usepackage{hyperref}
\usepackage[open]{bookmark}

\usepackage{txfonts}
\usepackage{color}
\usepackage{mathrsfs}

\newcommand{\sro}{Sr$_2$RuO$_4$}

\newcommand{\vk}{\vec{k}}

\newcommand{\kl}{\left(}
\newcommand{\kr}{\right)}

\newcommand{\ua}{\uparrow}
\newcommand{\da}{\downarrow}
\newcommand{\ksp}{_{\vec{k}}}

\newcommand{\e}{\epsilon}
\newcommand{\La}{\Lambda}

\newcommand{\kxy}{\kappa_{xy}}
\newcommand{\dxxyy}{$d_{x^2-y^2}+id_{xy}$}
\newcommand{\dzxzy}{$d_{zx}+id_{zy}$}
\newcommand{\pxy}{$p_x + ip_y$}

\begin{document}
\title{Spontaneous thermal Hall effect in three-dimensional chiral superconductors with gap nodes}

\author{Nobuyuki Yoshioka$^1$, Yoshiki Imai$^2$ and Manfred Sigrist$^3$}
\affiliation{$^1$Department of Physics, University of Tokyo, 7-3-1 Hongo, Bunkyo-ku, Tokyo 113-0033, Japan\\
$^2$Department of Applied Physics, Okayama University of Science, Okayama, Japan \\
$^3$ Institut f\"{u}r Theoretische Physik, ETH Z\"{u}rich, CH-8093 Z\"{u}rich, Switzerland}

\begin{abstract}
Generic chiral superconductors with three-dimensional electronic structure have nodal gaps and are not strictly topological. Nevertheless, they exhibit a spontaneous thermal Hall effect (THE), i.e. a transverse temperature gradient in response to a heat current even in the absence of an external magnetic field. While in some cases this THE can be quantized analogous to the Quantum Hall effect, this is not the case for nodal superconductors in general. In this study we determine the spontaneous THE for tight binding models with tetragonal and hexagonal crystal symmetry with chiral $p$- and $d$-wave superconducting phase. At the zero-temperature limit, the thermal Hall conductivity $ \kxy $ provides information on the structure of the gap function on the Fermi surface and the Andreev bound states on the surface. The temperature dependence at very low temperatures is determined by the types of gap nodes, point or line nodes, leading to characteristic power law behaviors in the temperature, as known for other quantities such as specific heat or London penetration depth. The generic behavior is discussed on simple models analytically, while the analysis of the tight-binding models is given numerically. 
\end{abstract}

\maketitle

\section{Introduction}\label{introduction}

Among all the unconventional superconducting phase, those which break time-reversal symmetry (TRS) belong to the most interesting cases as they combine two antagonistic features; superconductivity and magnetism. 
TRS breaking phases can generate intrinsic magnetism through spontaneous supercurrents at surfaces and intrinsic domain walls and around impurities or lattice defects \cite{volovik_1985,sigrist_1991}. 
Magnetic fields generated by such supercurrents have been observed by means of zero-field $\mu$SR \cite{luke_1998,biswas_2013}. One type of TRS violating superconductors are those which are chiral, or in the classification of Volovik and Gor'kov ''ferromagnetic'' \cite{volovik_1985}. 
These superconductors show topological properties and in particular produce subgap chiral quasiparticle state at surfaces with the normal vector perpendicular to the chiral axis, which is the orientation of the net angular momentum of the Cooper pairs \cite{volovik_1988}. 
Among the unconventional superconductors with broken TRS the most promising candidates for chiral Cooper pairing are UPt$_3$\cite{Joynt_2002,izawa_2014}, Sr$_2$RuO$_4$ \cite{luke_1998,mackenzie_2003}, URu$_2$Si$_2$\cite{kasahara_2007,goswami_2013} and SrPtAs\cite{biswas_2013,fisher_2014}. 

Sr$_2$RuO$_4$ represents the best-investigated case so far, but does not entirely conform with expectations. 
Early on, it has been guessed that this superconductor could be a so-called chiral $p$-wave superconductor of the type \pxy. 
While the $\mu$SR and polar Kerr effect measurements \cite{luke_1998,xia_2006} are fully compatible with such a chiral phase, no magnetic field due to currents of chiral edge states have been observed\cite{kirtley_2007,kallin_2009,curran_2014,mackenzie_2017, etter_2018}. 
This apparent absence of surface currents has put some doubts on the chiral nature of the superconducting phase in Sr$_2$RuO$_4$. However, it has to be noted that the charged supercurrent at the surface is not a universal topologically protected quantity, as the charge is not conserved in the coherent superconducting state \cite{huang_2014,bouhon_2014,huang_2015}. 
On the other hand, the Bogolyubov quasiparticles form topologically protected chiral Andreev bound states at the surface, due to bulk-edge correspondence, and carry energy, which is conserved. Therefore it was proposed that the analog of the Quantum Hall effect could be found in thermal transport through the thermal Hall effect (THE) or Righi-Leduc effect could provide a quantized topologically universal phenomenon \cite{read_2000,sumiyoshi_2013,imai_2016}. 

The thermal Hall conductivity is quantized in the case of a strictly topological superconducting phase which is fully gapped. 
In this case, we expect that in the low-temperature limit the thermal Hall conductivity $\kxy$ is expressed as
\begin{equation}
\kxy = \frac{\pi}{12} C T,
\end{equation}
where $ C $ is an integer, the Chern number, characterizing the chiral superconducting state.  Here, we set $k_B = \hbar = 1$ for simplicity. 
It is important to realize that this THE is spontaneous, i.e. no external field has to be applied. It is an intrinsic property of the superconductor. 
For Sr$_2$RuO$_4$, which has a quasi-two-dimensional band structure with three cylindrical Fermi surfaces\cite{mackenzie_2003}, this is in principle possible. However, most other superconductors are more three-dimensional and are most likely not truly topological, for the reason that they have zero nodes in the gap. Point nodes behave like Weyl points such that this type of superconductor behaves like a Weyl superconductor that generates Fermi arcs at certain surfaces \cite{meng_2012,fisher_2014,goswami_2013}. Despite not being topological in a strict sense, these superconductors nevertheless show a spontaneous THE which is not quantized.

In the present study, we describe the THE of such superconductors and determine which features are responsible for the magnitude of $ \kxy(T) $. We will show that
the zero-temperature limit is connected with Fermi arcs, while the temperature dependence at low temperatures shows power laws in $T$ due to low-energy quasiparticles around the gap nodes. For this purpose we will first  introduce tight-binding models on tetragonal and hexagonal crystal lattices 
for chiral pairing states of types, \pxy, \dzxzy and \dxxyy, for different types of Fermi surfaces. We then derive the basic formalism and analyze the generic properties of $ \kxy(T)$ for these cases in order to show which information can be extracted.

\section{Formulation of the model Hamiltonians}

We start our discussion introducing the general models for 3D chiral SC with nodal gaps in the quasiparticle spectrum.  For the derivation
of the Bogolyubov-de Gennes (BdG) mean-field Hamiltonian we consider a $d$-dimensional lattice version based on a tight-binding band structure 
and an attractive inter-site interaction responsible for Cooper pairing.  The Hamiltonian is composed of two terms
\begin{eqnarray}
H_{\mathrm{tot}}=H_{\mathrm{kin}} +H_{\mathrm{int}},
\end{eqnarray}
corresponding to the kinetic and the two-body interaction term, respectively, with
\begin{eqnarray}
H_{\mathrm{kin}} &=& - \sum_{i, j,s} t_{i,j} c_{i,s}^{\dagger} c_{j,s}+ h.c.,
\\
H_{\mathrm{int}}&=&\sum_{i,j,\alpha}V_{i,j}^{\alpha} \; n_{i,\alpha} n_{j,\alpha}.\label{gen_intham}
\end{eqnarray}
Here, $c^{(\dagger)}_{i,s}$ is the annihilation (creation) operator of an electron on the site $i$ with spin $s$ ($=\ua$ or $\da$) and $n_{i,\alpha} = \sum_{s,s'} c^{\dagger}_{i,s} \sigma_{s,s'}^{\alpha} c_{i,s'}$ is the charge ($\alpha=0$) and spin ($\alpha=1,2,3$) density operator on the site $i$ ($ \sigma^0 $ is the $2 \times 2 $-identity matrix and $ \sigma^{1,2,3} $ are the Pauli matrices). The hopping matrix elements between different sites $i$ and $j$ are denoted as $t_{i,j}$ and the chemical potential corresponds to $ t_{i,i} = \mu  $ in our notation.  In $H_{\mathrm{int}}$ we use a density-density type of interaction with coupling $V_{ij}^{\alpha}$, between the sites $ i$ and $j$, whereby we will neglect here onsite interaction ($i=j$) which would lead to conventional pairing. For spin-rotation symmetry we may impose the constraint $ V_{i,j}^{0} = V_{i,j}^{(c)} $ and $ V_{i,j}^{1,2,3} = V_{i,j}^{(s)} $. 

\subsection{Mean-field BdG Hamiltonian}

We can now turn to the usual mean-field decoupling leading to a BdG-type of Hamiltonian quadratic in the electron operators,
\begin{eqnarray}\label{generalham}
H_{\mathrm{mf}}=\frac{1}{2} \sum_{\vk} \Psi^{\dagger}_{\vk} \; H_{\mathrm{BdG}}(\vk) \; \Psi_{\vk},
\end{eqnarray}
where $\Psi_{\vk}^t = \left( c_{\vk,\ua}\ c_{\vk,\da}\ c^{\dagger}_{-\vk,\ua}\ c^{\dagger}_{-\vk,\da} \right)$ is the Nambu spinor with the electron operators in momentum space,
\begin{eqnarray}
c_{j,s} = \frac{1}{\sqrt{N}} \sum_{\vk} c_{\vk,s} e^{i \vk \cdot \vec{r}_j},
\end{eqnarray}
where $ N $ is the number of lattice sites and $ \vec{r}_j $ is the position of site $j$. This defines $H_{\mathrm{BdG}}(\vk)$ as a $4\times4$-matrix given by
\begin{eqnarray}
H_{\mathrm{BdG}}(\vk)&=&
\left( 
 \begin{array}{cc}
\hat{\epsilon}_{\vk}&\hat{\Delta}_{\vk}\\
\hat{\Delta}_{\vk}^{\dagger}&-\hat{\epsilon}_{-\vk}
 \end{array}
\right)
\end{eqnarray}
with
\begin{eqnarray}
\hat{\epsilon}_{\vk}
&=& \sigma^0 \e\ksp
\label{kin_mat},\\ \nonumber\\
\hat{\Delta}_{\vk}&=&
\left( 
 \begin{array}{cc}
\Delta^{\ua\ua}_{\vk}&\Delta^{\ua\da}_{\vk}\\
\Delta^{\da\ua}_{\vk}&\Delta^{\da\da}_{\vk}
 \end{array}
\right)
= i\sigma_y \left( \sigma^0 d_0(\vk)+\vec{\sigma}\cdot\vec{d}(\vk) \right).\label{int_mat}
\end{eqnarray}
Here, $\e\ksp$ is the tight-binding kinetic energy including the chemical potential,
\begin{eqnarray}
\e\ksp = - \sum_{\vec{R}_{i,j}} t_{i,j} e^{i \vk \cdot \vec{R}_{i,j}} ,
\end{eqnarray}
with $ \vec{R}_{i,j} = \vec{r}_j - \vec{r}_i $ being the lattice vector connecting site $ i $ and $j$. 
The components of the gap function, $\Delta^{s \tilde{s}}\ksp$, are defined through the mean field, first in real space
\begin{eqnarray}
\Delta_{j,i}^{s \tilde{s}} &=& - \sum_{\alpha, s',\tilde{s}'} V_{i,j}^{\alpha} \sigma_{s,s'}^{\alpha} \sigma_{\tilde{s}, \tilde{s}'}^{\alpha} \langle c_{j,\tilde{s}'} c_{i, s'} \rangle  \nonumber \\  
                                   &=& - \sum_{s',\tilde{s}'} \tilde{V}_{i,j}^{ss'\tilde{s}\tilde{s}'}   \langle c_{j,\tilde{s}'} c_{i,s} \rangle,
\end{eqnarray}
which is transformed in momentum space as
\begin{eqnarray}
\Delta^{s \tilde{s}}\ksp &=& \sum_{\vec{R}_{j,i}} \Delta_{j,i}^{s \tilde{s}} e^{-i \vk \cdot \vec{R}_{j,i}} = - \sum_{\vk', s', \tilde{s}'} \tilde{U}_{\vk, \vk'}^{ss'\tilde{s}\tilde{s}'} \langle c_{- \vk' \tilde{s}'} c_{\vk' s'} \rangle , \nonumber\\
\tilde{U}_{\vk, \vk'}^{ss'\tilde{s}\tilde{s}'} &=&  \sum_{\vec{R}_{j,i}}  \tilde{V}_{i,j}^{ss'\tilde{s}\tilde{s}'} e^{-i (\vk - \vk') \vec{R}_{j,i}},
\label{gap-k}
\end{eqnarray}
where $ \langle A \rangle $ denotes the thermal average of the operator $A$. The complex gap function is parametrized by $d_0 (\vk)$ and $\vec{d}(\vk) = \kl d_1(\vk),d_2(\vk),d_3(\vk) \kr$, corresponding to one spin-singlet- and three spin-triplet-components of the pairing state, respectively. 
In the following, we assume inversion symmetry such that parity is a good quantum number identifying the singlet configuration as the even-parity state ($ d_0(-\vk) = d_0(\vk) $) and the triplet components as the odd-parity ($ \vec{d} (-\vk) = - \vec{d} (\vk) $). 
Furthermore, we restrict to
unitary states, i.e., $\hat{\Delta}_{\vk} \hat{\Delta}_{\vk}^{\dagger}$ is proportional to $\sigma^0$ for all $ \vk $, avoiding effects of spin polarization\cite{sigrist_1991}.

For our study, it is sufficient to look at the cases with $ \vec{d} \parallel \hat{z} $ ($ d_1 (\vk) = d_2 (\vk) =0 $), yielding a separation of the $ H_{\mathrm{BdG}}(\vk) $-matrix into two $2\times 2 $-blocks within a two-dimensional version of the Nambu spinor $ {\tilde{\Psi}_{\vk}^t} = \left(c_{\vk,\uparrow}\ c^{\dagger}_{-\vk,\downarrow} \right)$. This leads to
\begin{eqnarray}
\tilde{H}_{\mathrm{mf}}&=&\frac{1}{2} \sum_{\vk} \tilde{\Psi}^{\dagger}_{\vk}\tilde{H}_{\mathrm{BdG}}(\vk)\tilde{\Psi}_{\vk},\\
\tilde{H}_{\mathrm{BdG}}(\vk)&=&\left( 
 \begin{array}{cc}
\epsilon_{\vk}&\Delta_{\vk}\\
\Delta^{*}_{\vk}&-\epsilon_{-\vk}
 \end{array}
\right),\label{kspace_bdg}
\end{eqnarray}
where $ \Delta_{\vk} = d_0 (\vk) $ for even-parity spin-singlet pairing and  $ \Delta_{\vk} = d_3 (\vk) $ for odd-parity spin-triplet pairing. The self-consistent equation now reads,
\begin{eqnarray}\label{selfconsistency}
\Delta_{\vk} = \frac{1}{N} \sum_{\vk'} U_{\vk \vk'} \frac{\Delta_{\vk'}}{2 E_{\vk'}} \tanh \left( \frac{E_{\vk'}}{2 T} \right)
\end{eqnarray}
where $ U_{\vk, \vk'} = U^{(0)}_{\vk,\vk'} - 3 U^{(s)}_{\vk, \vk'} $ for singlet and $ U_{\vk, \vk'} = U^{(0)}_{\vk, \vk'} + U^{(s)}_{\vk, \vk'} $ for triplet pairing. Moreover,
$ E_{\vk} = \sqrt{ \epsilon_{\vk}^2 + |\Delta_{\vk}|^2} $ is the Bogolyubov quasiparticle energy.

\subsection{Models of chiral superconductors}\label{MCS}
Now we consider concrete examples of chiral superconducting phases in three-dimensional systems with tetragonal and hexagonal crystal symmetry. The  corresponding point groups are $D_{4h}$ and $D_{6h}$, respectively. 
This choice is motivated by superconductors such as \sro, URu$_2$Si$_2$, UPt$_3$ and SrPtAs which are considered as candidates that realize chiral superconductivity. 
The possible chiral phases are listed in Table~\ref{models_table}, in which the states are classified according to their corresponding irreducible representation within the point groups. 
For the tetragonal system we consider both the spin-singlet and the spin-triplet phase, and for the hexagonal lattice we restrict to the spin-singlet case only. 
This allows us to cover most of the generic features, in particular the nodal gaps, which we would like to address in the following. 

\begin{table}[!htb]
\begin{center}\hspace{-2.5cm}\hspace{2.3cm}
\begin{tabular}{|c|ccc|} \hline
Lattice sym. & $D_{4h}$&  $D_{4h}$& $D_{6h}$\\
Irreducible rep. & E$_g$ & E$_u$ & E$_{2g}$ \\
Pairing & \dzxzy & \pxy & \dxxyy \\
Possible nodes & LN, PN($n = \pm1$) & PN($n=\pm1$) & PN($n=\pm1,\pm2$)\\
 \hline
\end{tabular}
\caption{The lattice symmetry, the irreducible representation of the point group, the pairing function of the Cooper pairs, possible nodes in the Fermi surface. LN and PN stand for the line node and point node, respectively. The topological charge of the nodal point is denoted by $n$. }
\label{models_table}
\end{center}
\end{table}

\begin{figure}[t]
\begin{center}
\begin{tabular}{c}
  \begin{minipage}{0.33\hsize}
    \begin{center}
     \resizebox{!}{0.85\hsize}{\includegraphics{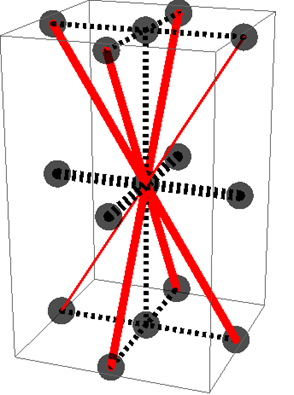}}
     \hspace{-2cm}\\ \ \ \ \ \ \ \ (a) $d_{zx}+id_{zy}$
   \end{center}
  \end{minipage}
  \begin{minipage}{0.33\hsize}
   \begin{center}
     \resizebox{!}{0.85\hsize}{\includegraphics{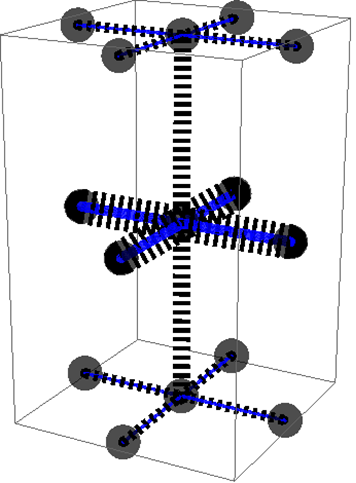}}
     \hspace{-2cm}\\ \ \ \ \ \ \ \ (b) $p_x+ip_y$
    \end{center}
  \end{minipage}
  \begin{minipage}{0.33\hsize}
   \begin{center}
     \resizebox{!}{0.85\hsize}{\includegraphics{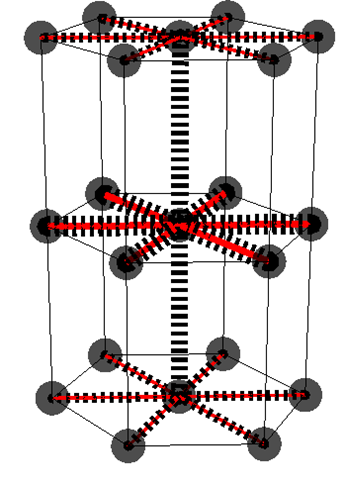}}
     \hspace{-2cm}\\ \ \ \ \ \ \   (c) $d_{x^2-y^2}+id_{xy}$
    \end{center}
  \end{minipage}
\end{tabular}
\end{center}
\caption{\label{interaction}The colored solid lines denote the pairing interaction, where  singlet (triplet) pairing is expressed by red (blue). Implicitly we assume specific crystal field which results in the corresponding irreducible representation: E$_g$ and E$_u$ of $D_{4h}$ and E$_{2g}$ of $D_{6h}$. The dashed lines indicate the nearest-neighbor hopping. }
\end{figure}

For the kinetic energy, it is sufficient to use a model with nearest-neighbor hopping. 
We consider a simple tetragonal lattice in the case of $ D_{4h} $,  and for $ D_{6h} $ simply stacking triangular lattices. 
We will investigate various shapes of Fermi surfaces which are accessible in these systems by merely changing the chemical potential. More complex shapes could be reached by introducing hopping integrals beyond nearest neighbors, which will, however, not provide further insights. 
The pairing interactions correspond to next-nearest-neighbor interaction for $ E_g $ in $ D_{4h} $, represented by $ \vec{R}_{i,j} = (\pm a,0,\pm c), (0, \pm a, \pm c) $, and to in-plane nearest-neighbor for $ E_u $ in $ D_{4h} $ with $ \vec{R}_{i,j} = (\pm a,0,0), (0, \pm a, 0)$ where $ a $ and $ c $ are the in-plane and out-of-plane lattice constants. 
For $ E_{2g} $ in $ D_{6h}$, only in-plane nearest-neighbor interaction is taken into account with $ \vec{R}_{i,j} = (\pm a,0,0), (\pm 1/2, \pm \sqrt{3}/2, 0), (\pm 1/2, \mp \sqrt{3}/2, 0)$. The geometry of pairings is depicted in Fig. \ref{interaction}. 

The tight-binding bands are given by 
\begin{eqnarray}
\e\ksp &=& -2t_1 (\cos k_x + \cos k_y ) -2t_2\cos k_z-\mu, \;  \mbox{for}  \;  D_{4h} \label{tb-d4h} \\
\e\ksp &=& - 2 t_1 \sum_{j=0}^2 \cos( \vk \cdot \vec{\delta_j}) -2t_2\cos k_z-\mu, \;  \mbox{for}  \;  D_{6h} \label{tb-d6h}
\end{eqnarray}
with $ t_1 $ and $ t_2 $ as hopping matrix elements. Without loss of generality, we set $a = c = 1$. The gap functions $ \Delta_{\vk} $ are obtained through Eq. (\ref{gap-k}), for $d_{zx} + id_{zy}$ belonging to the $E_g$ representation
in $ D_{4h} $,
\begin{eqnarray}
\Delta\ksp = \Delta\sin k_z(\sin k_x + i \sin k_y),  \label{tetrad_gap-d}
\end{eqnarray}
for $ p_x + i p_y $ in the $ E_u $ of $ D_{4h} $,
\begin{eqnarray} 
\Delta\ksp = \Delta(\sin k_x + i \sin k_y), \label{tetrad_gap-p}
\end{eqnarray}
and the $ d_{x^2-y^2}+id_{xy} $-state in $ E_{2g} $ of $ D_{6h} $,
\begin{eqnarray}
\Delta\ksp = \Delta\sum_{j=0}^2\omega^j \cos(\vec{k}\cdot\vec{\delta}_j) .\label{hexag_gap-d} 
\end{eqnarray}
Here $\omega = e^{2\pi i/3}$, and the direction unit vectors are defined as $\vec{\delta}_0 = (1,0,0), \vec{\delta}_1 = (1/2, \sqrt{3}/2,0), \vec{\delta}_2 = (-1/2, \sqrt{3}/2,0)$.
Like the tight-binding energy bands also the gap functions defined in this way are periodic in reciprocal lattice. All superconducting phase have their chiral axis along the $z$-direction.

\section{Topological properties}

In this section, we analyze the topological properties of our models. 
We introduce the Berry curvature in reciprocal space and the Chern number characterizing the chiral superconducting phases on the basis of the Berry connection. 
For our discussion, it turns out to be useful to define a Chern number by slicing the Brillouin zone. 
As all examples of our chiral phases have their chiral axis along $z$-direction, this is done by
considering a 2D Brillouin zone (BZ) (parallel to the $k_x$-$k_y$-plane) for a fixed $ k_z$. 
This type of Chern number can be determined in a rather simple way based on the Fermi surface topology and the presence of singular points of the gap function within the BZ, as we point out below. This scheme allows us also to determine the presence of the zero-energy Andreev bound states (ZEABS) on the surface BZ. 
The bulk-edge correspondence is clearly confirmed by comparing these ZEABS with the $k_z$-dependent Chern number.

\subsection{Chern number}\label{BC_introduction}
\begin{figure}[b]
\begin{center}
\begin{tabular}{c}
  \begin{minipage}{0.4\hsize}
    \begin{center}
     \resizebox{!}{\hsize}{\includegraphics{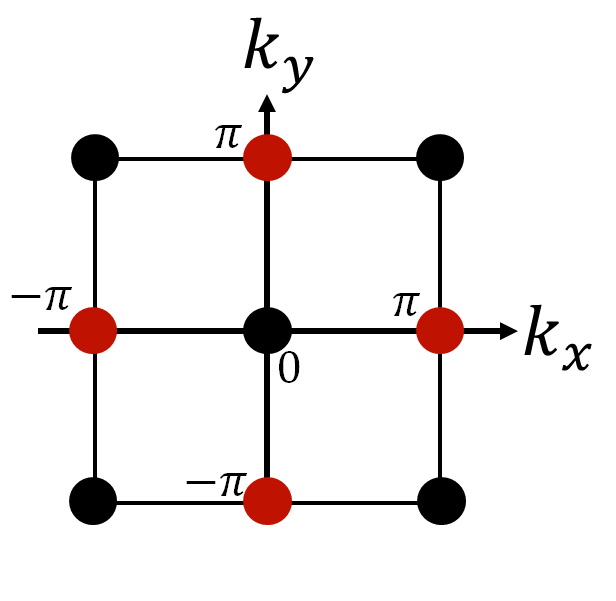}}
     \hspace{-2cm}\\ (a) 
   \end{center}
  \end{minipage}
  \begin{minipage}{0.3\hsize}
   \begin{center}
     \resizebox{!}{0.85\hsize}{\includegraphics{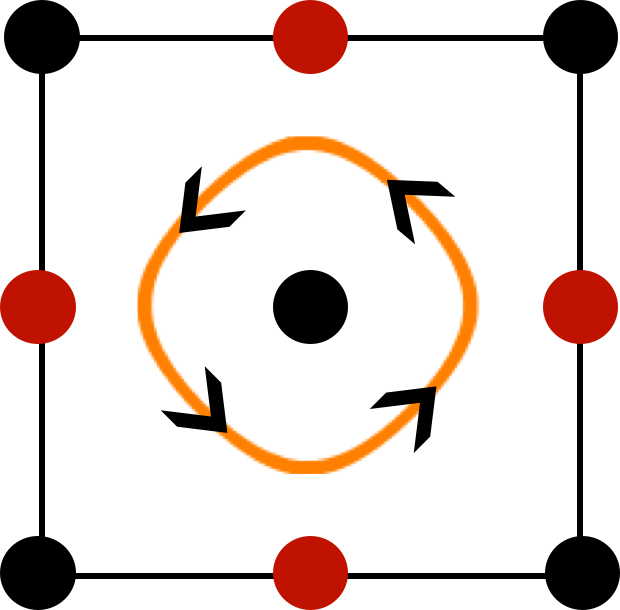}}
     \hspace{-2cm}\\ (b) 
    \end{center}
  \end{minipage}
  \begin{minipage}{0.3\hsize}
   \begin{center}
     \resizebox{!}{0.85\hsize}{\includegraphics{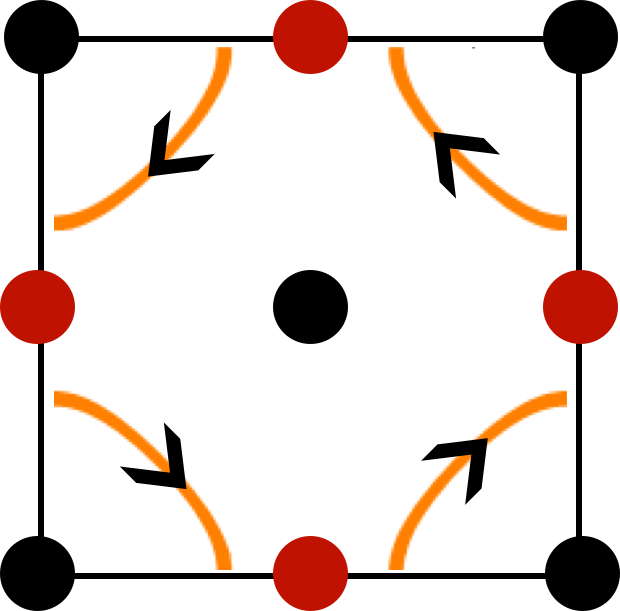}}
     \hspace{-2cm}\\  (c) 
    \end{center}
  \end{minipage}
\end{tabular}
\end{center}
\caption{\label{tetra_pndef} (a) The cross section of the Brillouin zone for the tetragonal lattice. The complex phase of the gap function winds $+(-)2\pi$ around the black (red) dots. (b)(c) Representative cross sections of the Fermi surface denoted by the orange line. The Chern number is given as $+1$ and $-1$, respectively.}
\end{figure}

Topological phases can be characterized by topological invariants. For time reversal symmetry breaking phases, the Chern number is the most common among them, and can be calculated through an integral of the Berry curvature in the reciprocal space. 
Generally, the Berry curvature $ \vec{\Omega}_{n,\vk} $ in the reciprocal space is a U(1)-gauge invariant quantity defined as the curl of the Berry connection $\vec{\mathcal{A}}_{n,\vk} $,
\begin{eqnarray}
\vec{\Omega}_{n,\vk} &=& \vec{\nabla}_{\vk}\times\vec{\mathcal{A}}_{n,\vk},\label{bc_eqn}\\
\vec{\mathcal{A}}_{n,\vk} &=& i\Braket{u_{n,\vk}|\vec{\nabla}_{\vk}|u_{n,\vk}},
\end{eqnarray}
where $\ket{u_{n,\vk}}$ is the $n$th eigenstate with energy $\epsilon_{n,\vk}$ of the Bloch Hamiltonian $H_{\vk}$ satisfying $H_{\vk}\ket{u_{n,\vk}} = \epsilon_{n,\vk}\ket{u_{n,\vk}}$. 

Considering our BdG Hamiltonian, only the $z$-component of the Berry curvature is relevant. Restricting here on a single electron band, we can express the Berry curvature as\cite{xiao_2010},
\begin{eqnarray}\label{berry_single}
\Omega^z_{\vk} =  \frac{1}{2}\hat{m}\ksp\cdot [\partial_{k_x}\hat{m}\ksp\times \partial_{k_y}\hat{m}\ksp]
\end{eqnarray}
with
\begin{eqnarray}
\vec{m}_{\vk} = \left(\mathrm{Im}\Delta\ksp,\mathrm{Re}\Delta\ksp,\e\ksp\right) \; 
\label{m-vec}
\end{eqnarray}
and $ \hat{m}_{\vk} = \vec{m}_{\vk} / |\vec{m}_{\vk}| $, as derived in  Appendix \ref{berrycal_app}. This allows us to define a topological invariant, a Chern number, for any fixed value of $ k_z $, 
\begin{eqnarray}\label{Nch_kz}
C(k_z) = 2\pi \int_{\mathrm{BZ (k_z) }}\frac{dk_xdk_y}{(2\pi)^2}\Omega^z_{\vec{k}},
\end{eqnarray}
where the integral only runs over the $k_x$-$k_y$-cross-section of the BZ for fixed $ k_z $ ($\mathrm{BZ}(k_z)$). This yields an integer value $ C(k_z)$ for any $ k_z $ where $ \Omega^z_{\vk} $ has no zeros in $\mathrm{BZ}(k_z)$. 

It turns out that this Chern number can be identified with the winding number of the gap function $ \Delta_{\vk} $ around the Fermi surface,
\begin{equation}
C(k_z) = \frac{1}{2 \pi} \oint_{{\rm FS}(k_z)} d \vk \cdot \vec{\nabla}_{\vk} {\rm arg} (\Delta_{\vk}).
\end{equation}
To use this convenient feature systematically, we setup the following scheme. Analyze the zeros of $ \Delta_{\vk} $ within the BZ and determine their ''charge'' or winding number $n $. For the example of the chiral $p$-wave state in Eq. (\ref{tetrad_gap-p}), we find one zero in the center and three at the boundary of BZ($k_z$) (see Fig. \ref{tetra_pndef}). Expanding $ \Delta_{\vk} = \Delta (\sin k_x + i \sin k_y) $ around these zeros we find,
\begin{eqnarray}
\vk_0 = (0,0,k_z) & \qquad & \Delta_{\vk} \approx + \Delta (q_x + i q_y) = + \Delta q e^{i \theta_{\vec{q}}} \\
\vk_1 = (\pi,\pi,k_z) & \qquad & \Delta_{\vk} \approx - \Delta (q_x + i q_y) = - \Delta q e^{i \theta_{\vec{q}}}  \\
\vk_2 = (\pi,0,k_z) & \qquad & \Delta_{\vk} \approx - \Delta (q_x - i q_y) = - \Delta q e^{-i \theta_{\vec{q}}} \\
\vk_3 = (0,\pi,k_z) & \qquad & \Delta_{\vk} \approx + \Delta (q_x - i q_y) = + \Delta q e^{-i \theta_{\vec{q}}}  
\end{eqnarray}
where $ \vec{q} = \vk - \vk_n $ and $ \theta_{\vec{q}} $ is the polar angle of $ (q_x,q_y) $ and $ q = |\vec{q}| $. The charge $ n $ is defined through $ e^{i n \theta_{\vec{q}}} $.
In Fig. \ref{tetra_pndef}(a) the black dots have $ n=+1 $ and the red dots $ n=-1 $. The Chern number is determined by the cross section of the Fermi surface at given $ k_z $. Following the Fermi surface line in positive direction, the Chern number $ C(k_z)$ is given by the total charge encircled. 
This is $+1$ in Fig. \ref{tetra_pndef}(b) for the electron-like Fermi surface around the BZ center and $ -1 $ for the hole-like Fermi surface,  which surrounds the corner point with charge $ +1 $ in opposite orientation (orientation of the arrows), in Fig. \ref{tetra_pndef}(c). Note that we have discussed  the state with positive chirality as given in Eq. (\ref{tetrad_gap-p}).  The charges as well as the Chern number change sign, if we consider the the operation of time reversal that yields opposite chirality. 

\begin{figure}[t]
\begin{center}
\begin{tabular}{c}
  \begin{minipage}{0.4\hsize}
    \begin{center}
     \resizebox{!}{\hsize}{\includegraphics{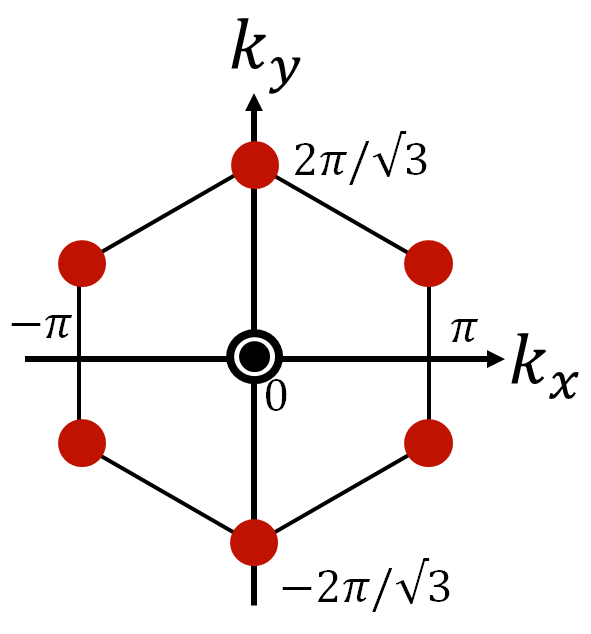}}
     \hspace{-2cm}\\ (a) 
   \end{center}
  \end{minipage}
  \begin{minipage}{0.3\hsize}
   \begin{center}
     \resizebox{!}{0.85\hsize}{\includegraphics{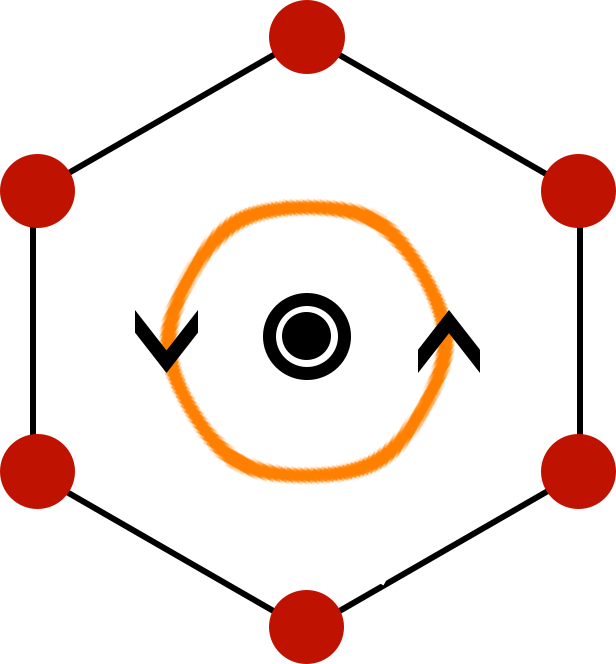}}
     \hspace{-2cm}\\ (b) 
    \end{center}
  \end{minipage}
  \begin{minipage}{0.3\hsize}
   \begin{center}
     \resizebox{!}{0.85\hsize}{\includegraphics{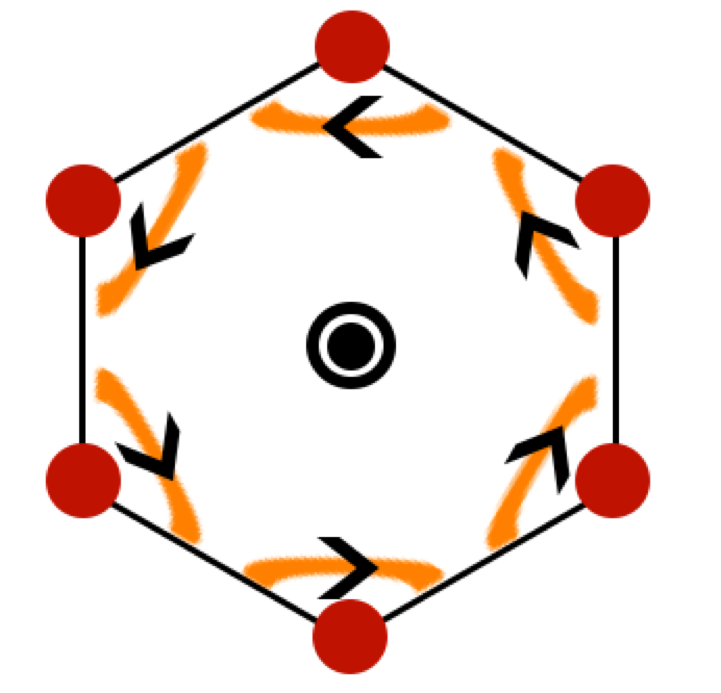}}
     \hspace{-2cm}\\    (c) 
    \end{center}
  \end{minipage}
\end{tabular}
\end{center}
\caption{\label{hexa_pndef} (a) The cross section of the Brillouin zone for the hexagonal lattice. The complex phase of the gap function winds $+4\pi$ around the black double circle. (b)(c) A cross section of the Fermi surface denoted by the orange line. The Chern number is $+2$ in the either case.}
\end{figure}

The analogous picture we obtain for the chiral $ d $-wave phase with $ \Delta_{\vk} $ in Eq. (\ref{hexag_gap-d}), which has the zeros at (see Fig. \ref{hexa_pndef}(a)),
\begin{eqnarray}
\vk_0 = (0,0,k_z) & \qquad & \Delta_{\vk} \approx  \Delta (q_x + i q_y)^2 =  \Delta q^2 e^{i 2 \theta_{\vec{q}}}, \\
\vk_1 = (0,4 \pi/\sqrt{3},k_z) & \qquad & \Delta_{\vk} \approx  \Delta (q_x - i q_y) = \Delta q e^{-i \theta_{\vec{q}}}, \\
\vk_2 = (2\pi, 2\pi/\sqrt{3},k_z) & \qquad & \Delta_{\vk} \approx  \Delta (q_x - i q_y) = \Delta q e^{-i \theta_{\vec{q}}}, 
\end{eqnarray}
yields charge $ +2 $ for the gap zero at the BZ center (black double dot) and $-1$ at the BZ corners (red dots). This leads to a Chern 
number $ +2 $ for the electron-like Fermi surface in Fig. \ref{hexa_pndef}(b) 
and $-2$ for the hole-like Fermi surface of Fig. \ref{hexa_pndef}(c), as two zeros are encircled. 

The Chern number $ C(k_z) $ is a piece-wise constant function of $k_z$. As $k_z$ varies, $ C(k_z)$ shows jumps from one integer to another, corresponding to a ''topological phase transition'' where the gap on the Fermi surface cross-section vanishes, as in a usual topological transition.

\begin{figure}[b]
\begin{center}
\begin{tabular}{l}
    \begin{minipage}{0.45\hsize}
    \begin{center}
     \resizebox{!}{0.9\hsize}{\includegraphics{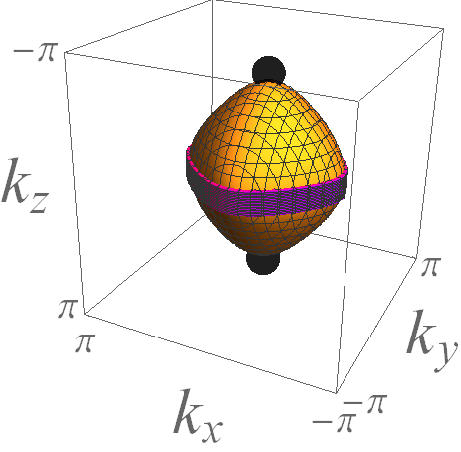}}
     \hspace{-2cm}\\ (a) Type A FS
   \end{center}
  \end{minipage}
  \begin{minipage}{0.5\hsize}
   \begin{center}
        \resizebox{!}{0.8\hsize}{\includegraphics{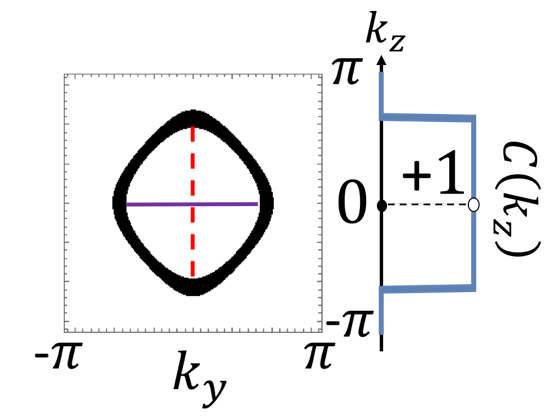}}
     \\ (d) Type A SBZ and $C(k_z)$
    \end{center}
  \end{minipage}       
  \\ 
   \begin{minipage}{0.45\hsize}
    \begin{center}
     \resizebox{!}{0.9\hsize}{\includegraphics{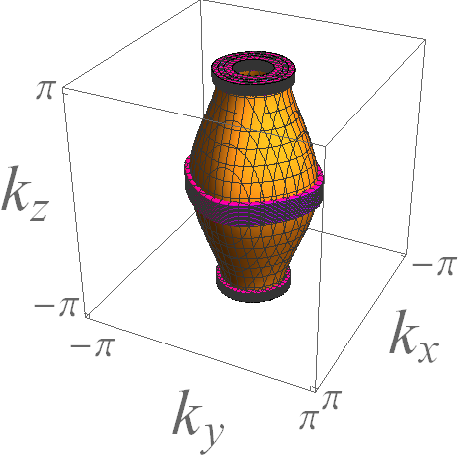}}
     \hspace{-2cm}\\ (b) Type B FS
   \end{center}
  \end{minipage}
  \begin{minipage}{0.5\hsize}
   \begin{center}
        \resizebox{!}{0.8\hsize}{\includegraphics{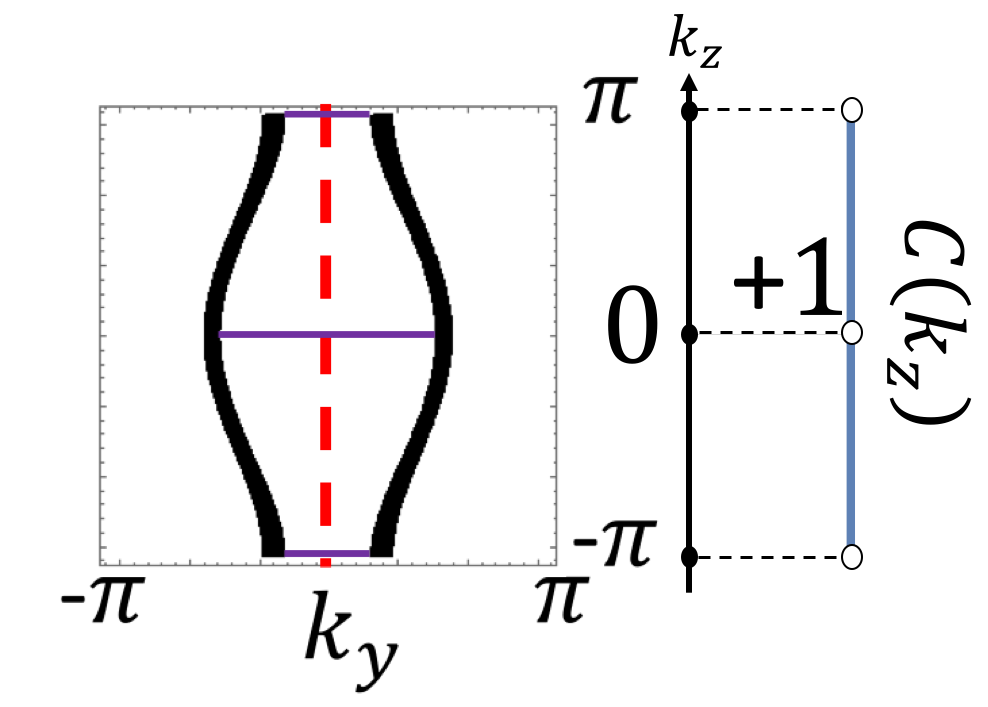}}
     \hspace{-2cm}\\ (e) Type B SBZ and $C(k_z)$
    \end{center}
  \end{minipage}
  \\ 
  
   \begin{minipage}{0.45\hsize}
   \begin{center}
     \resizebox{!}{0.9\hsize}{\includegraphics{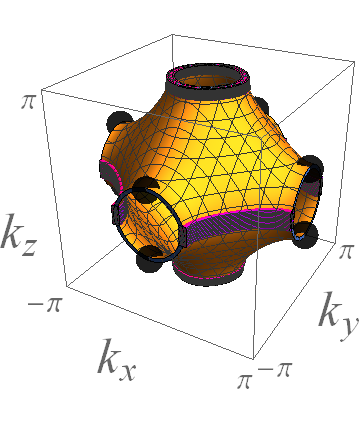}}
     \hspace{-3cm}\\  (c) Type C FS
    \end{center}
  \end{minipage}
    \begin{minipage}{0.5\hsize}
   \begin{center}
     \resizebox{!}{0.8\hsize}{\includegraphics{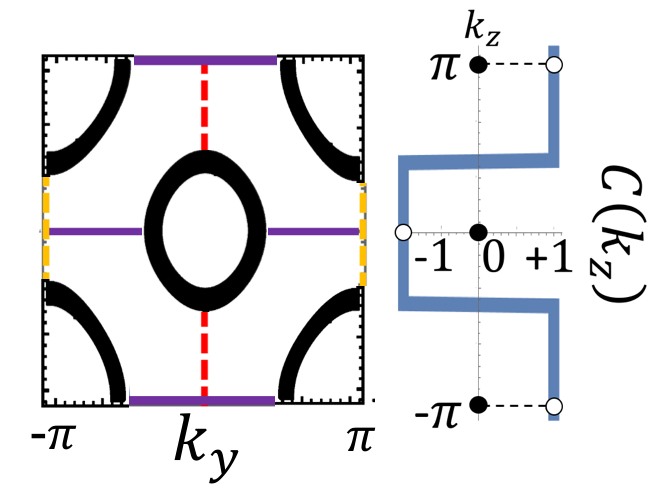}}
     \hspace{-2cm}\\   (f) Type C SBZ and $C(k_z)$
    \end{center}
  \end{minipage}
  
\end{tabular}
\end{center}
\caption{\label{tetra_fermi} 
\textcolor{black}{
 (a)(b)(c)The representative Fermi surfaces of \dzxzy-wave superconductor on a tetragonal lattice. While nodal lines are located at $k_z=0,\pm\pi$, nodal points are (a)  at ($k_x,k_y$)=(0,0) (b) absent, (c) at the sides of the Brillouin zone. The parameters are taken as ($t_2/t_1,\mu/t_1$) = (0.85,-3.1), (0.4,-2.6),  and (1, -1), respectively. (d)(e)(f) The (100) surface Brillouin zone and the $k_z$-dependence of the Chern number. The red (yellow) dotted lines denote positive (negative) chiral edge states, and the purple real lines correspond to bulk-gap closing due to the nodal line.  
 }
 }
\end{figure}

\begin{figure}[t]
\begin{center}
\begin{tabular}{l}
    \begin{minipage}{0.45\hsize}
    \begin{center}
     \resizebox{!}{0.9\hsize}{\includegraphics{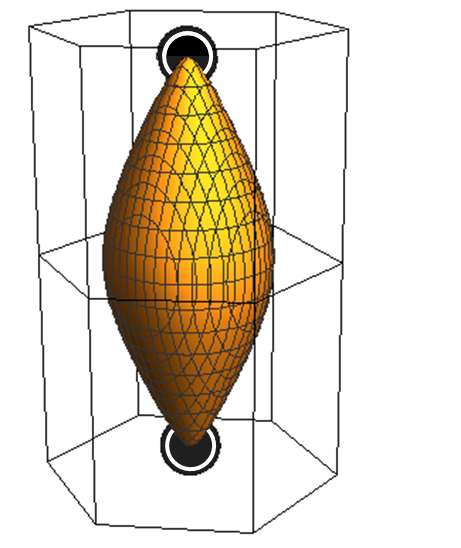}}
     \hspace{-2cm}\\ (a) Type A FS
   \end{center}
  \end{minipage}
  \begin{minipage}{0.5\hsize}
   \begin{center}
        \resizebox{!}{0.8\hsize}{\includegraphics{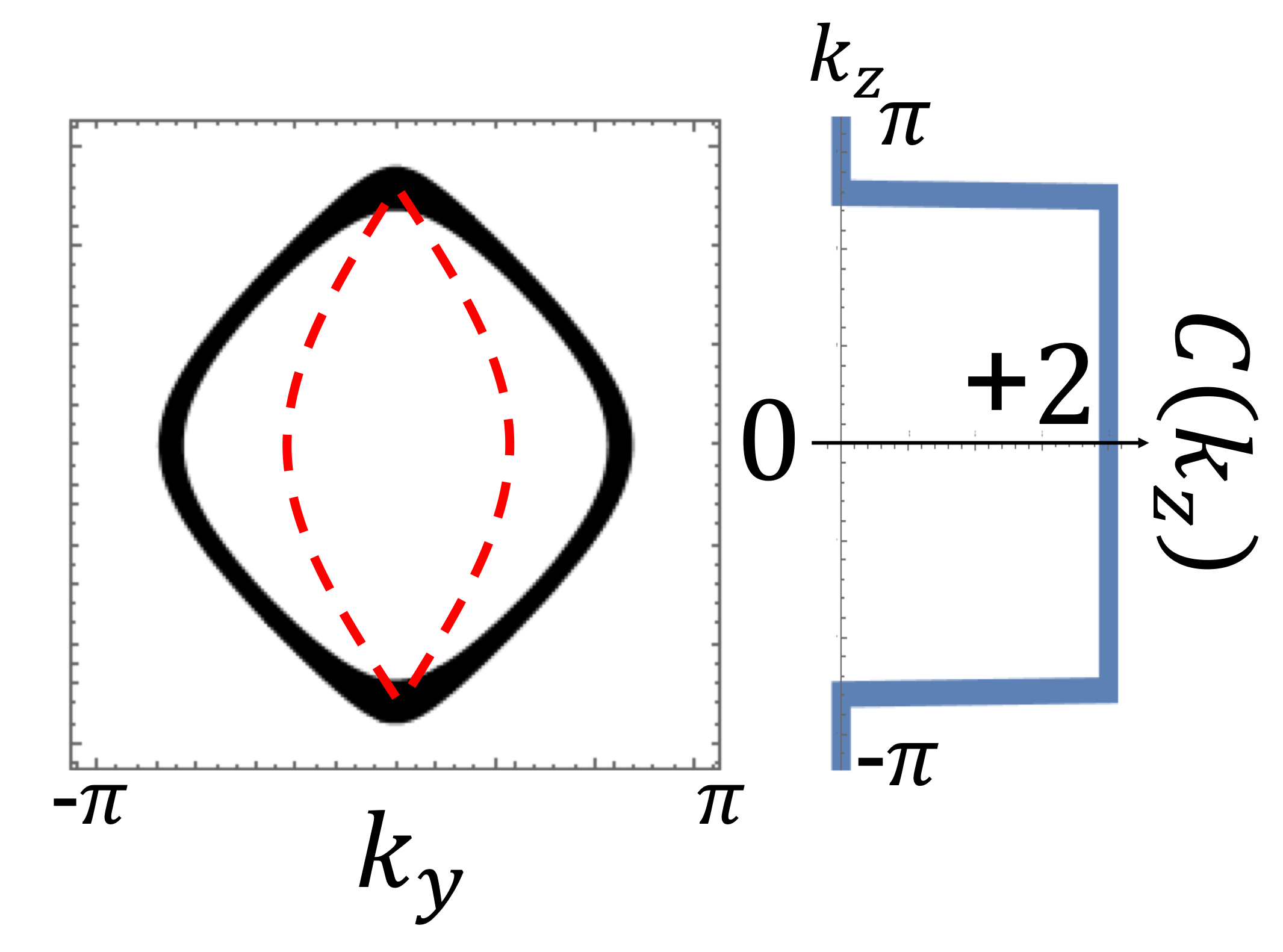}}
     \\ (d) Type A SBZ and $C(k_z)$
    \end{center}
  \end{minipage}       
  \\ 
   \begin{minipage}{0.45\hsize}
    \begin{center}
     \resizebox{!}{0.9\hsize}{\includegraphics{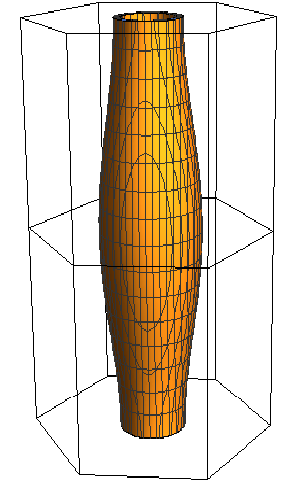}}
     \hspace{-2cm}\\ (b) Type B FS
   \end{center}
  \end{minipage}
  \begin{minipage}{0.5\hsize}
   \begin{center}
        \resizebox{!}{0.8\hsize}{\includegraphics{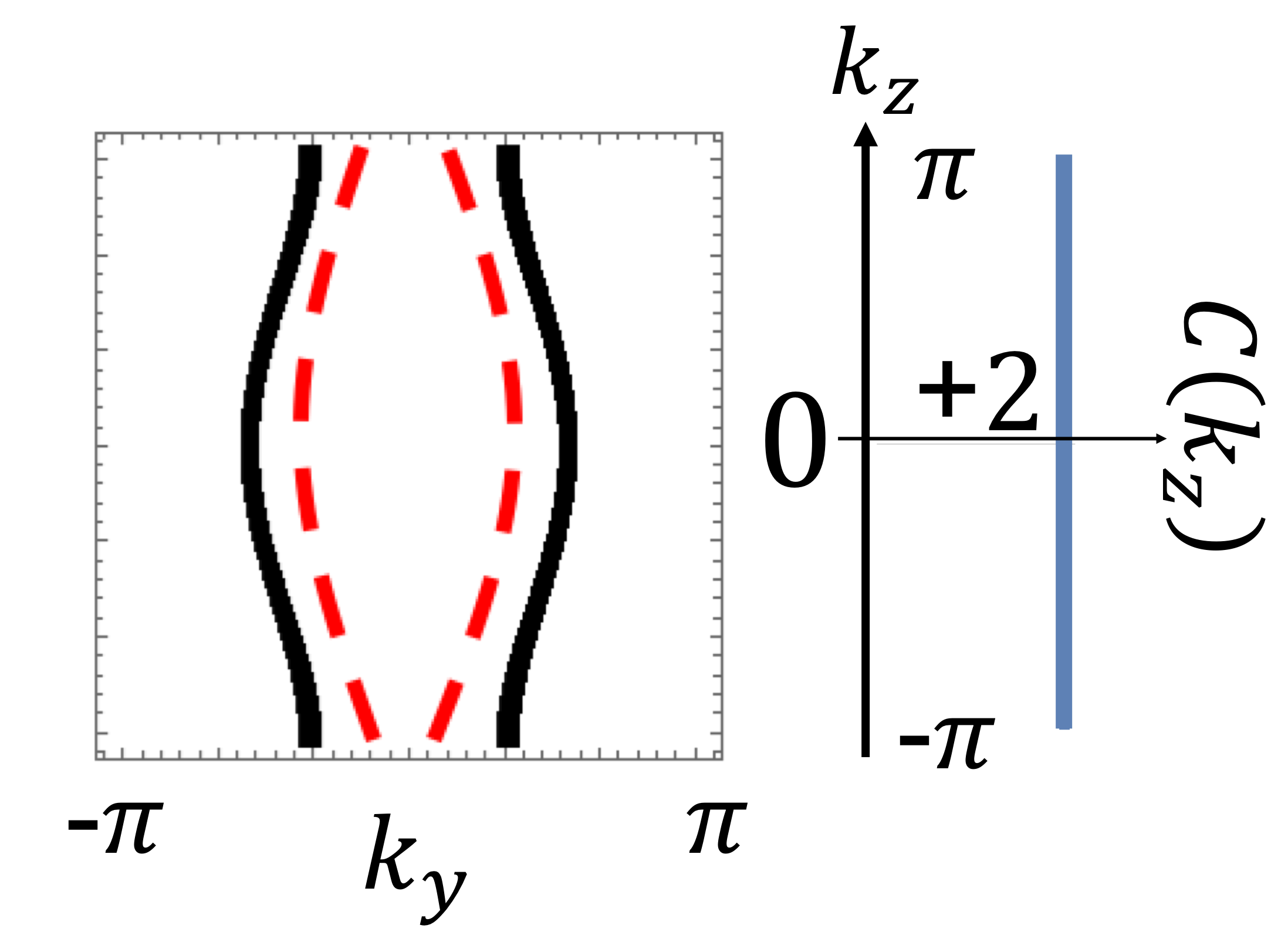}}
     \hspace{-2cm}\\ (e) Type B SBZ and $C(k_z)$
    \end{center}
  \end{minipage}
  \\ 

  \begin{minipage}{0.45\hsize}
   \begin{center}
     \resizebox{!}{0.9\hsize}{\includegraphics{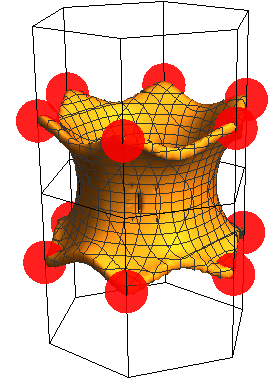}}
     \hspace{-3cm}\\  (c) Type C FS
    \end{center}
  \end{minipage}
    \begin{minipage}{0.5\hsize}
   \begin{center}
     \resizebox{!}{0.8\hsize}{\includegraphics{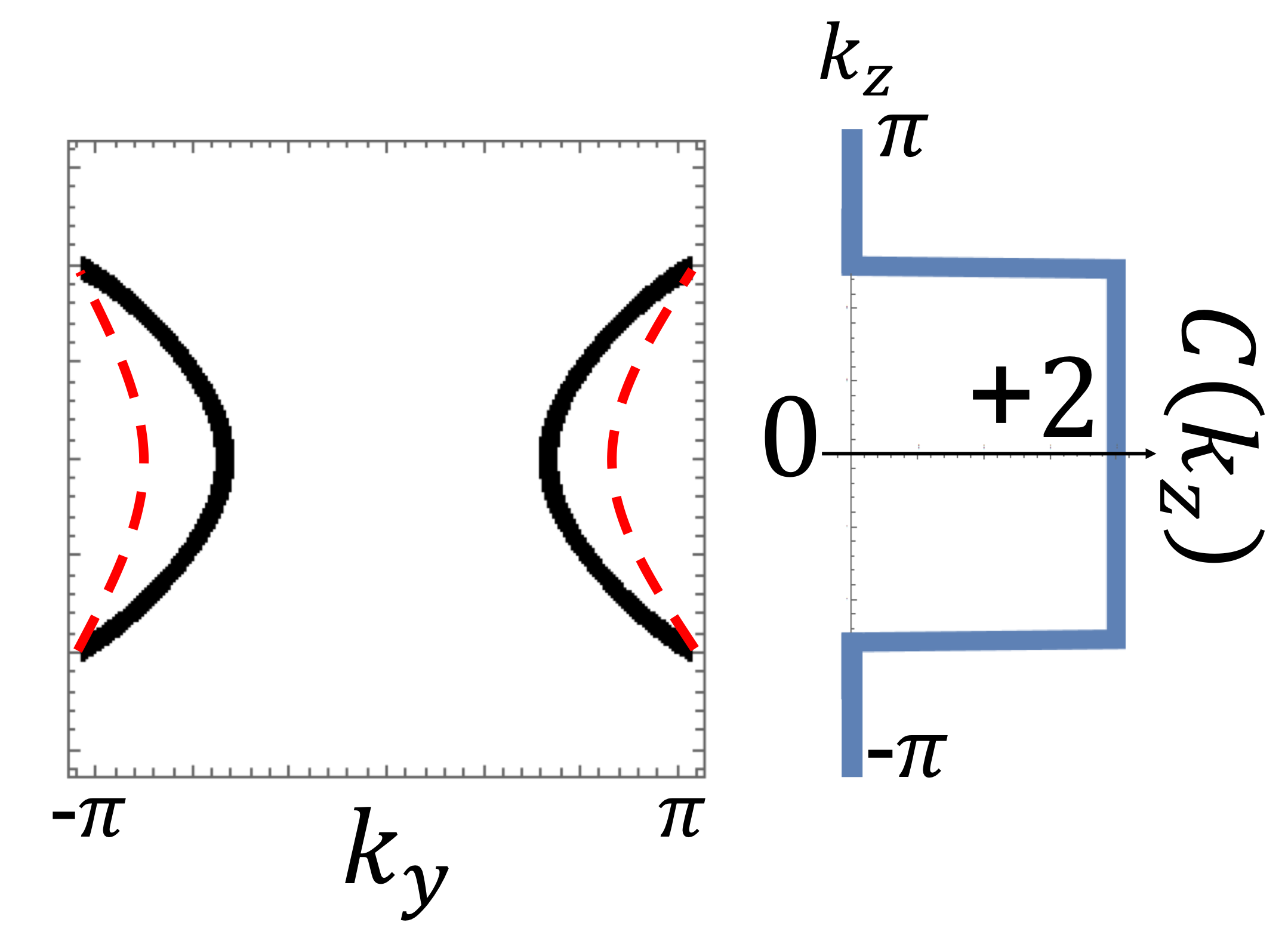}}
     \hspace{-2cm}\\  (f) Type C SBZ and $C(k_z)$
    \end{center}
  \end{minipage}

\end{tabular}
\end{center}
\caption{\label{hexa_fermi}  
\textcolor{black}{
(a)(b)(c)  The Fermi surface of \dxxyy-wave superconductor on a hexagonal lattice. Nodal points are (a) at ($k_x,k_y$)=(0,0), (b) absent, (c) at the corners. The parameters are taken as ($t_2/t_1,\mu/t_1$) =$(1.5,-3.5)$, $(0.5,-3.5)$, and $(-1.5,3.5)$, respectively. (d)(e)(f) The (100) surface Brillouin zone and the $k_z$-dependence of the Chern number. The red lines denote positive chiral edge states. 
}
}

\end{figure}

\subsection{Bulk-edge correspondence}

The Chern number $ C(k_z)$ is directly reflected in the presence and structure of edge states through the bulk-edge correspondence. 
Here, we illustrate this for a few examples, assuming a planar sample surface for our superconductor with its normal vector along the $x$-direction and translational symmetry for the other two directions. We focous on the ZEABS which belong to the chiral edge modes of Bogolyubov quasiparticle states propagating in a positive or negative direction along the $y$-direction depending on the associated Chern number. Again we always assume, for simplicity,  a SC state with positive chirality, defined by the sign of the charge at $ k_x = k_y = 0 $.


{\it Tetragonal symmetry:} As a first concrete case, we consider the  \dzxzy-wave SC for three different shapes of the FS as shown in Fig. \ref{tetra_fermi}, which can be obtained by different choices of the hopping matrix elements and the band filling. Fig. \ref{tetra_fermi}(a) shows a type A FS which is electron-like and closed. The gap function has a horizontal line node at $ k_z = 0 $ (purple line) and point nodes at $ \vec{k} = (0,0,\pm k_0) $ (black dot corresponding to black dots in Fig. \ref{tetra_pndef}), the top and bottom of the FS ($ k_0$ depends on the band filling). In Fig. \ref{tetra_fermi}(d) we see the Chern number $ C(k_z) $ which is $ +1 $ for $ -k_0 < k_z < k_0 $ and zero otherwise. The corresponding ZEABS, or ``the Fermi surface arc'', 
is marked by a dashed red line for $ k_y =0 $ and $ -k_0 < k_z < k_0 $. 
Here, we find one ZEABS with positive chirality in the range of $ k_z $ where $ C(k_z) = + 1 $. Generally, the number of chiral ZEABS with positive and negative chirality, denoted by $N_+(k_z)$ and $N_-(k_z)$, respectively, obeys the relation\cite{hatsugai_1993},
\begin{equation}
N_+(k_z)-N_-(k_z) = C(k_z) .
\end{equation}
Stepping to Fig. \ref{tetra_fermi} (b) with a type B FS which open along the $k_z$-axis, we find line nodes for $ k_z = 0 $ and $ \pm \pi $. Here $ C(k_z) = +1 $ for all $ k_z $ in the BZ as shown in Fig. \ref{tetra_fermi}(e) with a chiral ZEABS for $k_y=0$ along $k_z$-axis (dashed red line). 

More complex is the type C FS shown in Fig. \ref{tetra_fermi}(c), which is open along all main axes. Again we find nodal lines in the gap for $ k_z = 0 $ and $ \pm \pi $ as the case of the type B FS. Additionally there are point nodes at BZ boundary at $ \vec{k} = (0,\pm \pi, \pm k_0') $ and $ (\pm \pi,0, \pm k_0') $. By scanning $k_z$, we find  the Chern number is $ C(k_z) = -1 $ for $ - k_0' < k_z < + k_0' $, where FS($k_z $) is hole-like, and $ C(k_z) = + 1 $ for $ k_0' < |k_z| < \pi $, where FS($k_z $) is electron-like. The corresponding ZEABS are drawn as Fermi arcs in Fig. \ref{tetra_fermi}(f) by a dashed red line (positive chirality mode) at $ k_y =0$ in the range of $ C(k_z) = +1 $ and by a dashed yellow line (negative chirality mode) at $ k_y = \pm \pi $ where  $ C(k_z) = -1 $. 
Note, to be precise, that $ C(k_z) $ is not defined for $ k_z $ where we have a horizontal line node. 

The characterization of the FS can be done in the same way for the \pxy-wave case in a system with tetragonal symmetry. 
Here, the main difference compared to the  \dzxzy-wave SC is the absence of the horizontal line nodes in the gap, which is insignificant since line nodes has no influence on the topological structure of the state, i.e., there is no topological transition if $k_z$ passes a line node. 
Considering the same FS structures, we find the same Chern numbers, apart from the singularities for certain $ k_z $ with the line nodes.

{\it Hexagonal symmetry:} In Fig. \ref{hexa_fermi}, some representative cases for a \dxxyy-wave SC on a hexagonal lattice are shown. 
The most notable difference between the previous two models, apart from  the underlying lattice structure, is the higher orbital angular momentum of the Cooper pairs.  
The component $L_z$, the projection on the $z$-axis, is $\pm2$ and hence the Chern number at a given $k_z$ can take on values $\pm2$ \cite{schnyder_2015, volovik_1988}. 
The labeling of the FS types is analogous to tetragonal lattice. The type A FS hosts two point nodes at $ \vec{k} = (0,0,\pm k_0) $ (marked by double black points in Fig. \ref{hexa_fermi}(a)) with charge +2 as in Fig. \ref{hexa_pndef}. For the range $ | k_z | < k_0 $, we find $ C(k_z) = + 2 $ leading ot two ZEABS connecting the points $ k_z = \pm k_0 $ ($k_y =0 $) where the nodal points are the source of two Fermi arcs, in accordance with the bulk-edge correspondence (Fig. \ref{hexa_fermi}(d)). 

For the open type B FS as  in Fig. \ref{hexa_fermi}(b), we find the Chern number $ C(k_z) = + 2 $ for all $ k_z $ in the BZ  and correspondingly two ZEABS of positive chirality as depicted in Fig. \ref{hexa_fermi}(e). The case of type C FS in Fig. \ref{hexa_fermi}(c) has nodal points with charge $-1$ at the corners of the BZ boundary, colored red following Fig. \ref{hexa_pndef}, for the range $ |k_z| < k_0' $ where the Chern number is $ C(k_z) = + 2 $. This yields two ZEABS with positive chirality, connecting the projection of the nodal points. 

For all these case, the bulk-edge correspondence is clearly obeyed.


\section{Thermal Hall conductivity}

We turn now to the THE or Righi-Leduc effect of 3D chiral SCs
with the aim to extract information connected with topology of the pairing state and the gap nodes. 
In a chiral SC, the THE appears spontaneously in the SC phase without applying a magnetic field. 
The expression of the thermal Hall conductivity is given by \cite{luttinger_1964, matsumoto_prl_2011,matsumoto_prb_2011, qin_2011,sumiyoshi_2013}
\begin{eqnarray}\label{kappa_TSC}
\kappa_{xy}&=&\frac{1}{4\pi T}\int_{-\infty}^{\infty}d\epsilon \epsilon^2\kl -\frac{df}{d\epsilon}\kr \La(\epsilon),\\
\La(\epsilon) &=& 2\pi \int_{BZ}\frac{d^3 k}{(2\pi)^3}\sum_{n}\Omega^z_{n,\vk}\Theta(\epsilon-E_{n,\vk}),
\end{eqnarray}
where $f(\epsilon)=1/(1+\exp(\epsilon/T)$) is the Fermi-Dirac distribution, $\Theta(x)$ is the step function, and $n $ is the band index corresponding to the $n$-th energy band. 
In the low-temperature limit ($ T \ll T_c $), we find the approximate expression,
\begin{eqnarray}
\kappa_{xy}\sim\kxy^L&=&\frac{\La(0)}{4\pi T}\int_{-\infty}^{\infty}d\epsilon \epsilon^2\kl -\frac{df}{d\epsilon}\kr, \label{approx-L}\\
\La(0) &=& 2\pi \int_{BZ}\frac{d^3 k}{(2\pi)^3}{\sum_{n}}'\Omega^z_{n,\vk}.\label{La0}
\end{eqnarray}
Note that the sum is taken over bands $ n $ with negative energy only ($ E_{n, \vk} < 0 $), corresponding to the occupied bands. 

\subsection{The leading order}
As the leading order in the low-temperature limit, we obtain the following $T$-linear term,
\begin{equation}
\kappa_{xy}^L = \frac{\La(0)}{2}\ \frac{\pi T}{6}.
\label{low-t-kap}
\end{equation}
To determine $ \La(0) $, we treat the $ k_z $-integration separately from the other two dimensions, and use Eq.(\ref{Nch_kz})\cite{sumiyoshi_2013}
to rewrite Eq.(\ref{La0}), 
\begin{equation}
\La(0) = \int_{-\pi}^{+\pi} \frac{dk_z}{2 \pi} {\sum_{n}}'  \int_{BZ(k_z)} \frac{d^2 k}{2\pi}\Omega^z_{n,\vk} = \int_{-\pi}^{+\pi} \frac{dk_z}{2 \pi} {\sum_{n}}' C_n(k_z) .
\end{equation}


Let us consider here the three types of FS for systems with tetragonal crystal symmetry in the \dzxzy-wave state, which is
shown in Fig. \ref{tetra_fermi}. Among these, only the type B FS takes an integer value, $ \La(0) = 1 $, corresponding effectively to a two-dimensional system. 
We find $  \La(0) =k_0 / \pi $ for the type-A FS,  and 
\begin{equation}
\La(0) = \frac{\pi - k_0'}{\pi} - \frac{k_0'}{\pi} = \frac{\pi - 2 k_0'}{\pi},
\end{equation}
for the type-C FS.
In these latter two cases, $ \La(0) $ is not an integer in general, as $ C(k_z)$ is not constant as function of $ k_z $, but only contributes in the range where we have ZEABS, or the Fermi arcs, in the surface BZ. In this way, $ \La(0) $ contains the information on the ''length'' of the Fermi arcs shown in Fig. \ref{tetra_fermi}(d)-(f), or more accurately, the distance between the projections of the Weyl points connected by the Fermi arc.

\subsection{Corrections due to nodal gaps}
The limiting behavior  in Eq. (\ref{low-t-kap}), which is connected with the topological properties, that results from the approximation Eq. (\ref{approx-L})  is valid only in the very low-temperature limit. 
Correction to $\kxy$  under increasing temperatures provide signatures of the quasiparticle excitations which are dominated by the gapless excitations originating from the existence of line and point nodes in the gap. It is helpful to consider the structure of $\La(\epsilon)$ which determines the low-$T$ behavior of $\kxy$. For nodal gaps, one finds that the leading correction for small $ \epsilon $ ($ | \epsilon | \ll \Delta $) takes the form of a power law,
\begin{equation}\label{Lambda_expand}
\La(\epsilon) \approx \La(0) + M_p \left|\epsilon \right|^p,
\end{equation}
where $ M_p $ is a coefficient depending on details of the model, while the power $ p $ is an integer and uniquely determined by the topology and shape of the nodes. From Eq. (\ref{Lambda_expand}), we straightforwardly  obtain the leading correction to $ \kxy^{L} $,
\begin{eqnarray}
\Delta\kxy&\sim&\frac{M_p}{4\pi}T^{p+1}\int d\epsilon |\epsilon|^{p+2}\kl-\frac{d}{d\epsilon}f(\epsilon)  \kr\\
               &=&\frac{M_p}{4\pi} \frac{(p+2)!(2^{p+1}-1)}{2^p}\zeta(p+2) T^{p+1},
\end{eqnarray}
where $\zeta(s)$ is the Riemann zeta function.

\begin{table}[b]
\begin{center}
\hspace{-2cm}

\begin{tabular}{|c|c|} \hline
type of nodes &$\Delta\kxy$\\ \hline \hline
line & $\propto T^2$\\ \hline
\kern-\tabcolsep\begin{tabular}{c}
point
 \\
(winding = $n$)
\end{tabular}
&$\propto T^{1+2/|n|}$ 
\\ 
 \hline
\end{tabular}
\caption{The temperature dependence of the correction terms originating in nodes. }
\label{LNandPN}
\end{center}

\begin{center}
\begin{tabular}{|c|l|c|c|} \hline
Pairing&FS type&$\Delta\kxy$ & dominant node \\ \hline \hline
\dzxzy  
 &Fig. \ref{tetra_fermi}: A, B, C & $ \propto T^2$ & line \\ \hline
\pxy &Fig. \ref{tetra_fermi}: B & $\propto e^{-\Delta/T}$ & nodeless \\ 
 &Fig. \ref{tetra_fermi}: A,C & $ \propto T^3$ & point $ n=1 $ \\ \hline
\dxxyy &Fig. \ref{hexa_fermi}: B &$\propto e^{-\Delta/T}$ & nodeless \\ 
 &Fig. \ref{hexa_fermi}: A & $\propto T^2$ & point $ n=2 $ \\ 
 &Fig. \ref{hexa_fermi} C & $\propto T^3$ & point $ n=1 $ \\  \hline
\end{tabular}
\caption{The temperature dependence of correction terms in different pairing function and the Fermi surface. }
\label{summary_models}
\end{center}
\end{table}
In the following, we explore how the nodal structure of the gap function is reflected in the power $p$ both analytically and numerically.
The effect of nodes and the results in the three concrete models discussed above are summarized in 
Table \ref{LNandPN} and \ref{summary_models}, respectively.

\subsubsection{Effect of nodal lines}


We use a simple example to calculate the corrections to $\kxy^L$ when line nodes dominate the low-energy excitations.
Assuming $d_{zx} + id_{zy}$-wave pairing on an isotropic Fermi surface, we consider the following band and gap structure,
\begin{eqnarray}
\e\ksp &=& A(\vec{k}^2-k_F^2), \\
\Delta\ksp &=& \frac{\Delta_0}{k_F^2} k_z(k_x +ik_y),
\end{eqnarray}
where $A = \hbar^2/2m^*$ with $ m^* $ the effective mass, $\Delta_0$ is the gap magnitude, and $k_F$ is the Fermi wave length. For this closed FS, we safely ignore the periodic structure of the spectrum in reciprocal space. 
A nodal line is realized at $k_z=0$, the equator of the Fermi sphere. We then focus on the quasiparticle spectrum 
in the vicinity of this line node,
\begin{equation}
E_{\vec{k}}^2 =  \e\ksp^2+|\Delta\ksp|^2 \approx 4A^2 k_F^2 k^2  + \Delta_0^2 \frac{k_z^2}{k_F^2},
\end{equation}
where $k$ measures the deviation of the wave vector from the FS in the $k_x$-$k_y$-plane. Now we determine the
Berry curvature from 
\begin{equation}
\vec{m}_{\vec{k}} = \left(\Delta_0 \frac{k_z k_y}{k_F^2} , \Delta_0 \frac{k_z k_x}{k_F^2}, A(\vec{k}^2-k_F^2) \right).
\end{equation}
With the approximations near the nodal line, we obtain
\begin{equation}
\Omega^z_{\vec{k}} =  \frac{A \Delta_0^2 k_z^2}{k_F^2 E_{\vec{k}}^3}.
\end{equation}
The integration of the Berry curvature is straightforward,
\begin{eqnarray} 
\La(\e) &=& \La(0) + \int \frac{d^3k}{(2 \pi)^2} \Omega^z_{\vec{k}} \Theta(\e - E_{\vec{k}}) \nonumber \\
& \approx & \La(0) - M_{\mathrm{LN}} | \e | + O(\e^2),\label{approxLN}
\end{eqnarray}
where
\begin{equation}
M_{\mathrm{LN}} = \frac{k_F}{8  \Delta_0}
\end{equation}
and $ \La(0) = k_F/\pi $.
Using Eq. (\ref{approxLN}), the thermal Hall conductivity can be obtained as
\begin{eqnarray}\label{kxy_LN}
\kxy =
\frac{\pi \La(0)}{12}T - \frac{9\zeta(3)}{4 \pi}M_{\rm LN}T^2 + O\kl T^3 \kr.
\end{eqnarray}
Here we restrict to the leading contribution of the line nodes and neglect the effect of the two point nodes along $ k_z $, which yield a contribution of higher power in $T$. 
Our analysis can easily be extended to the system with multiple nodal lines. The second term of Eq. (\ref{kxy_LN}) is then replaced by a sum of different contributions.

\begin{figure}[t]
 \begin{center}
 \begin{tabular}{c}
\hspace{-0.5cm}
  \begin{minipage}{0.45\hsize}
    \begin{center}
     \resizebox{!}{0.9\hsize}{\includegraphics{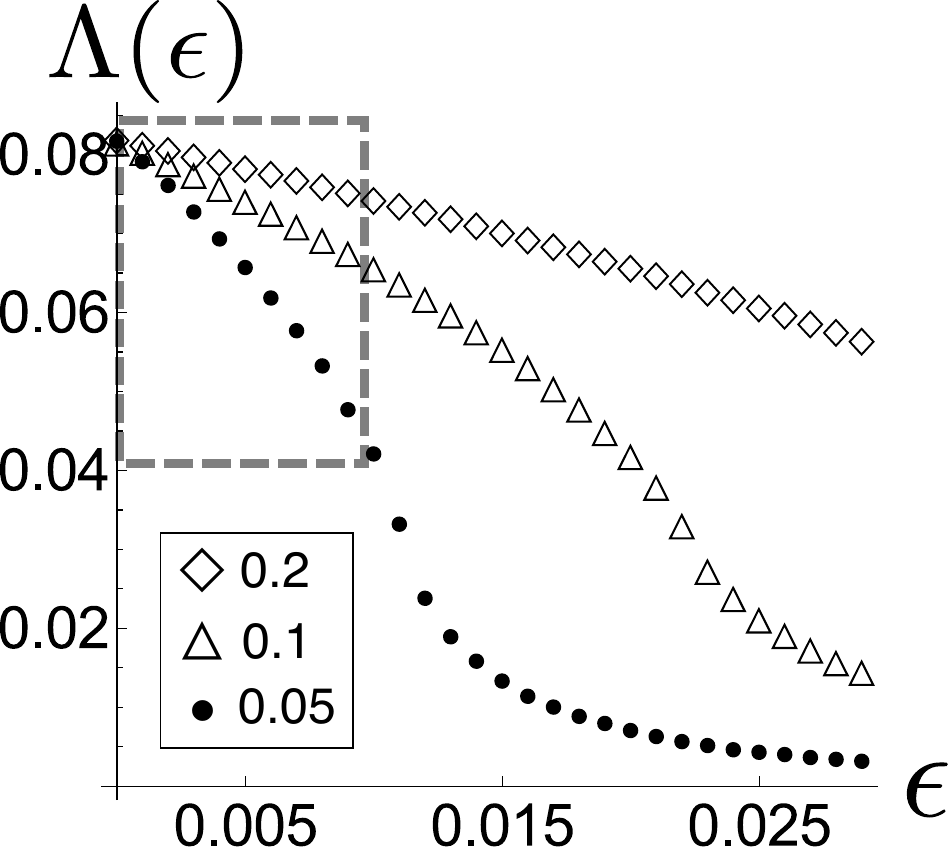}}\\
   \ \ \ \ \ \ \ \ \ \ \ \ \   (a) 
   \end{center}
  \end{minipage}
   \hspace{0.3cm}
  \begin{minipage}{0.45\hsize}
   \begin{center}
     \resizebox{!}{0.9\hsize}{\includegraphics{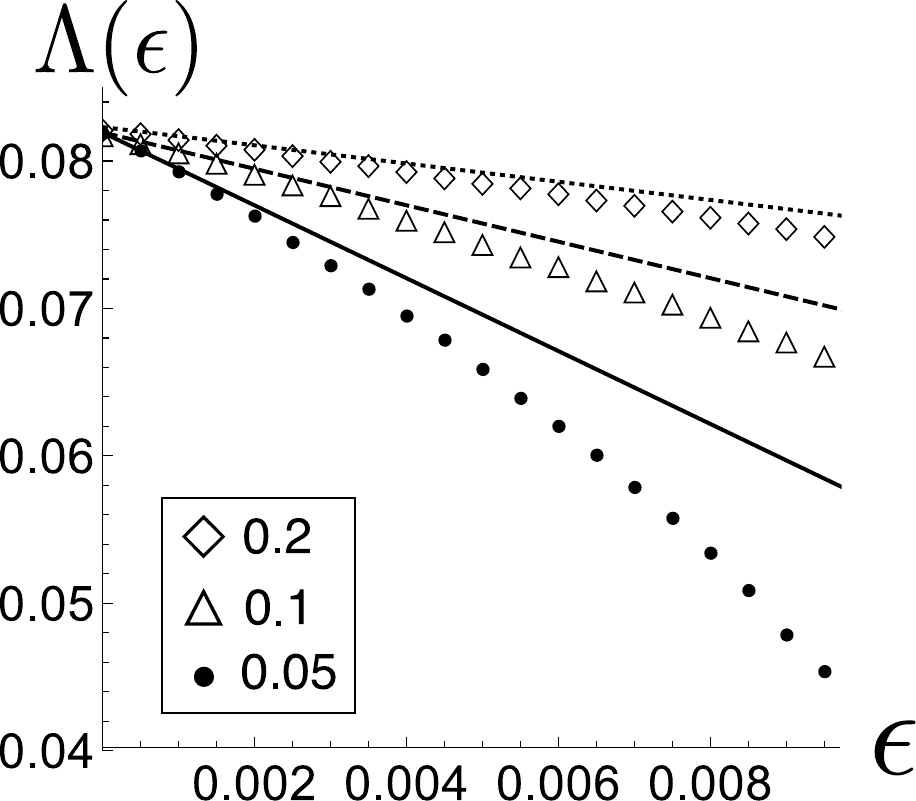}}\\
\ \ \ \ \ \ \ \ \ \ \ \ \   (b) 
    \end{center}
  \end{minipage}
 \end{tabular}
    \caption{\label{tetrad_A_lambda} (a) The $\e$ dependence of $\La$ for \dzxzy-wave SC with A-type FS for model of Eq. (\ref{tb-d4h}) with the parameters $t_2/t_1 = 3, \mu/t_1 = -9.8$, and the unit for the energy, $\epsilon$, is $t_1$, which is taken as unity. The filled circle, open triangle, and open diamond are numerical results for $\Delta/k_F^n = 0.05, 0.1, 0.2$, respectively, for the gap function of Eq. (\ref{tetrad_gap-d}). (b) Enlarged region marked by gray dotted rectangle in (a). The solid, dashed, and dotted lines correspond to the fit using Eq. (\ref{approxLN}).}
 \end{center}
\end{figure}

In Fig. \ref{tetrad_A_lambda}, we display the numerical result of $\La(\e)$ in \dzxzy-wave SC for a type-A FS with a single line node and two point nodes of $ n=1 $. The line nodes dominate the low-energy part of $\La(\e)$ and the Fermi arc has a finite length so that  $ \La(0) < 1 $.  
The linear lines are fits by  Eq. (\ref{approxLN}), where we determine $ k_F $ from Eq. (\ref{tb-d4h}) at the nodal points, i.e.  $-4t_1 -2t_2\cos(k_F) - \mu = 0$, and approximate $\Delta_0$ by  $\Delta$.   It fits the low-energy behavior of $\La(\e)$ well, supporting the validity of our analytical calculation. In particular, also the inverse-$\Delta_0 $-dependence is reproduced well. Obviously $ \La(\epsilon) $ has to be stronger suppressed by smaller $ \Delta_0 $, which is a measure for the strength of the superconducting phase, as the normal electron density is detrimental to the thermal Hall effect. 

\begin{figure}[t]
 \begin{center}
 \begin{tabular}{c}
  \begin{minipage}{0.45\hsize}
    \begin{center}
     \resizebox{!}{0.9\hsize}{\includegraphics{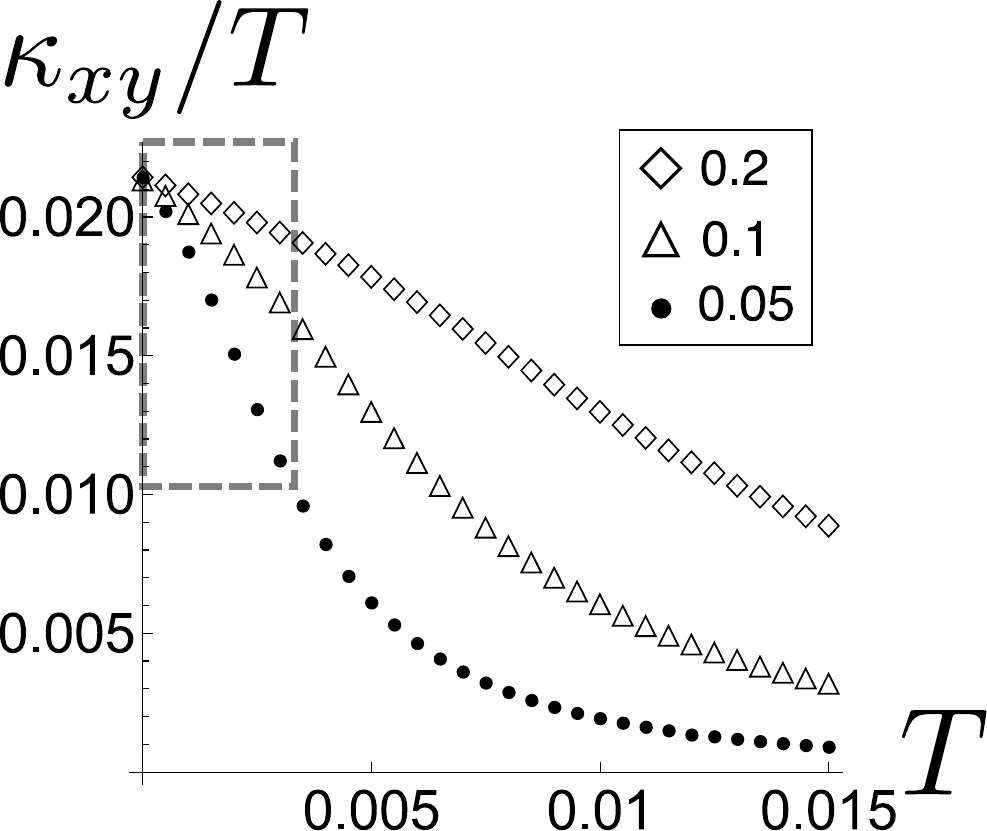}}\\   
       \ \ \ \ \ \ \ \ \ \ \ \ \   (a)    
       \end{center}
   \end{minipage}
      \hspace{0.4cm}
   \begin{minipage}{0.45\hsize}
   \begin{center}
           \resizebox{!}{0.9\hsize}{\includegraphics{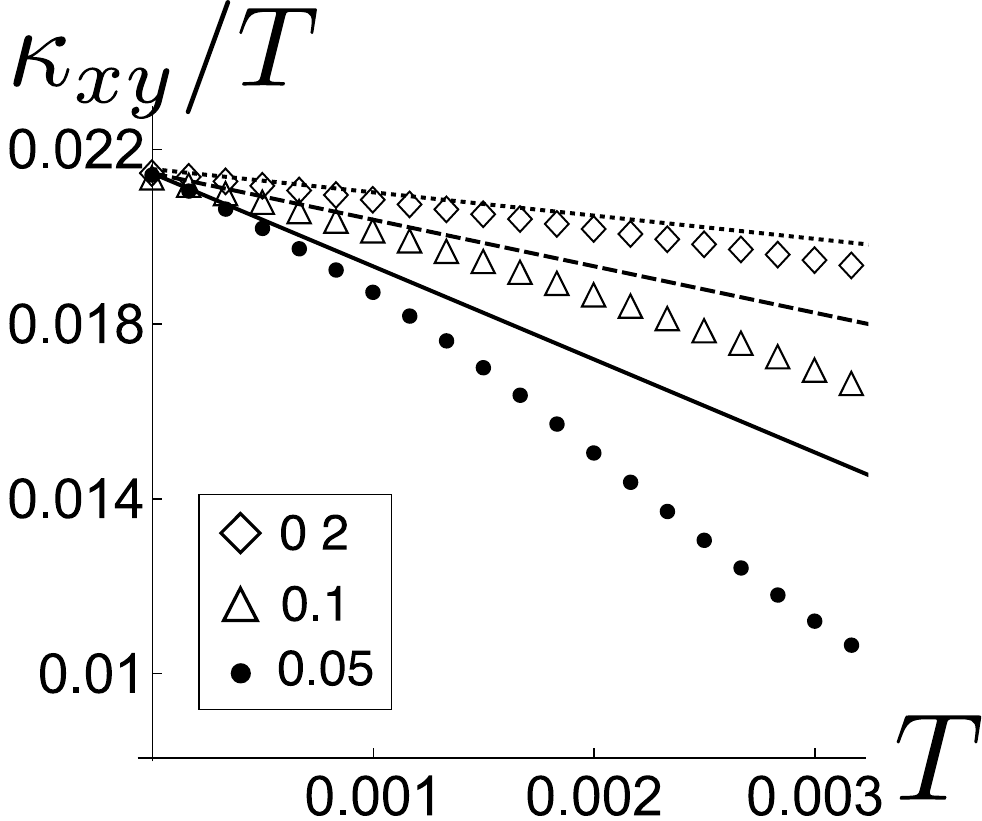}}
           \ \ \ \ \ \ \ \ \ \ \ \ \   (b) 
   \end{center}
  \end{minipage}
 \end{tabular}
    \caption{\label{tetrad_A_kxy} (a) $\kappa_{xy}/T$ numerically calculated for \dzxzy-wave SC in a tetragonal lattice with a FS type A with the parameters identical to Fig. \ref{tetrad_A_lambda}. 
    {The unit for the temperature, $T$,  is taken as $t_1 = 1$. }
    (b) Enlarged region marked by gray dotted rectangle in (a). The solid, dashed, and dotted lines correspond to the fit using Eq. (\ref{kxy_LN}).}
 \end{center}
\end{figure}

In Fig. \ref{tetrad_A_kxy}, we show the corresponding temperature dependence of $ \kxy $. We observe a linear approach to the zero-temperature value of $ \kxy/T $. This limiting value provides information of topological nature on the Fermi arc and the linear slope is a consequence of the line node. 
The lines in Fig. \ref{tetrad_A_kxy} (b) are fits by Eq. (\ref{kxy_LN}).


\subsection{Effect of nodal points}\label{effect_PN}

Next, we turn to the influence of point nodes by considering the examples of the \pxy-wave and the \dxxyy-wave superconducting phases, both with point nodes along the $k_z$-axis. However, it is important to notice that the winding number of the point nodes for type-A Fermi surfaces is different for the two cases; $ n =1 $ for \pxy-wave and $ n= 2 $ for \dxxyy-wave. This is reflected in the $ k $-dependence of $ \Delta_{\vk} $ around the nodes. 
For our analytical treatment, we consider again a spherical
FS and the corresponding gap functions,
\begin{eqnarray}
\e\ksp &=&A (\vec{k}^2-k_F^2) , \\
\Delta\ksp &=& \frac{\Delta_0}{k_F^n} (k_x +ik_y)^n .
\end{eqnarray}
We concentrate on the point node at $ \vk=(0,0,k_F) $ where the Bogolyubov quasiparticle spectrum can be approximated by
\begin{equation}
E_{\vk}^2 \approx A^2 (k^2+2k_Fq)^2 + \Delta_0^2 \frac{k^{2n}}{k_F^{2n}}
\end{equation}
with $ k_z = k_F + q $ and $ k^2 = k_x^2 + k_y^2 $. Following Eq. (\ref{m-vec}) we write
\begin{equation}
\vec{m}^{(1)} = \left(\Delta_0 \frac{k_y}{k_F} , \Delta_0 \frac{k_x}{k_F}, A(\vk^2-k_F^2) \right)
\end{equation} 
for $ n =1 $ and
\begin{equation}
\vec{m}^{(2)} = \left(\Delta_0 \frac{2k_x k_y}{k_F^2} , \Delta_0 \frac{k_x^2-k_y^2}{k_F^2}, A(\vk^2-k_F^2) \right)
\end{equation}
for $ n =2 $. From this we obtain the approximate contribution of the point nodes to the Berry curvature
\begin{equation}
\Omega_{\vk}^{(1)} \approx \frac{A \Delta_0^2}{2 k_F^2 E_{\vk}^3} (k^2 - 2 k_F q)
\end{equation}
and
\begin{equation}
\Omega_{\vk}^{(2)} \approx -\frac{4 A \Delta_0^2}{ k_F E_{\vk}^3} k^2q .
\end{equation}
Analogous to the previous section, we can now calculate the deviation of $ \Lambda(\epsilon) $ from $ \Lambda(0) $ due to the given point node as
\begin{eqnarray}
\Lambda^{(n)}(\epsilon)-\Lambda^{(n)}(0) &=&  \int \frac{d^3k}{(2 \pi)^2} \Omega_{\vk}^{(n)} \Theta(\epsilon-E_{\vk})  \nonumber \\
&\approx&  M_{\rm PN}^{(n)} |\epsilon|^{n/2} + O(\epsilon^{1+n/2}) .
\label{lambda-pn}
\end{eqnarray}
with $ \Lambda^{(n)}(0) = n k_F/\pi $ where $ n $ counts the number of Fermi arcs of length $k_F/\pi $ between the projected Weyl points. 
The evaluation of the integral leads to 
\begin{equation}
M_{\rm PN}^{(n)} = \begin{cases} \frac{k_F}{6 \pi \Delta_0^2} , & n=1, \\ \frac{k_F}{8 \Delta_0,} & n=2,\end{cases}
\end{equation}
where the point node at $ \vk = (0,0,-k_F) $ would give the same contribution. Consequently, the leading correction in the temperature dependence 
has the form
\begin{equation}
\kappa_{xy}^{(1)} = \frac{\pi \Lambda^{(1)}(0)}{12} T - \frac{7 \pi^3}{30} M_{\rm PN}^{(1)} T^3 
\label{kxy-pn1}
\end{equation}
and
\begin{equation}
\kappa_{xy}^{(2)} = \frac{\pi \Lambda^{(2)}(0)}{12} T - \frac{9 \zeta(3)}{2 \pi} M_{\rm PN}^{(2)} T^2 .
\label{kxy-pn2}
\end{equation}

\begin{figure}[t]
 \begin{center}
 \begin{tabular}{c}
  \begin{minipage}{0.45\hsize}
    \begin{center}
     \resizebox{!}{0.9\hsize}{\includegraphics{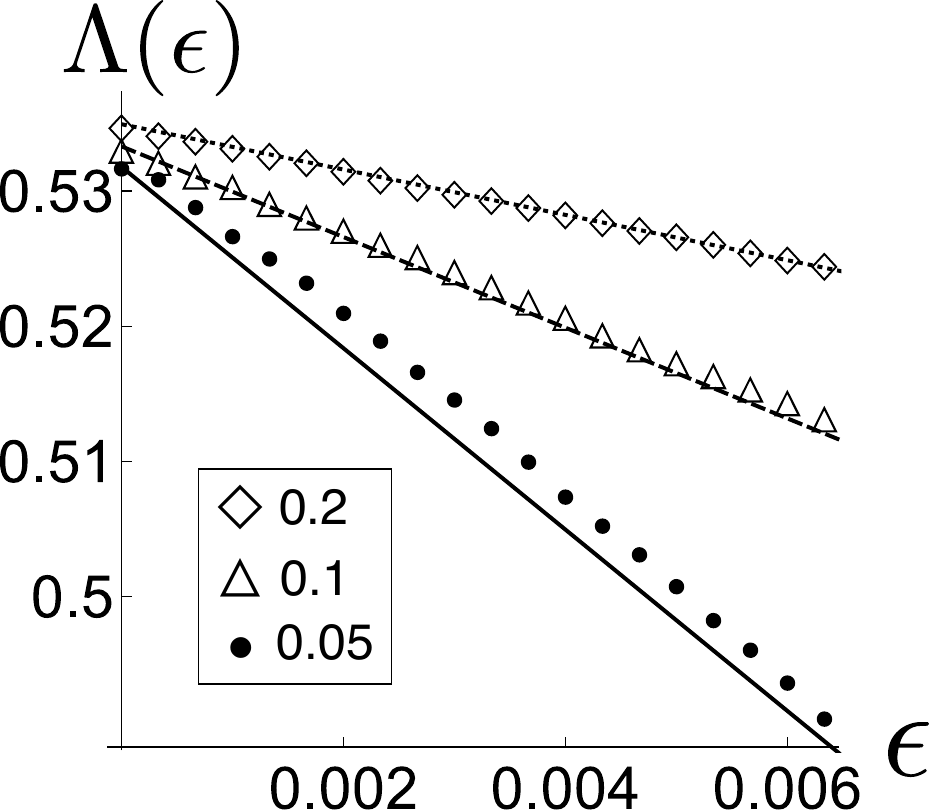}}\\
      (a)
     \end{center}
     \end{minipage}
     
     \begin{minipage}{0.45\hsize}
     \begin{center}
           \resizebox{!}{0.9\hsize}{\includegraphics{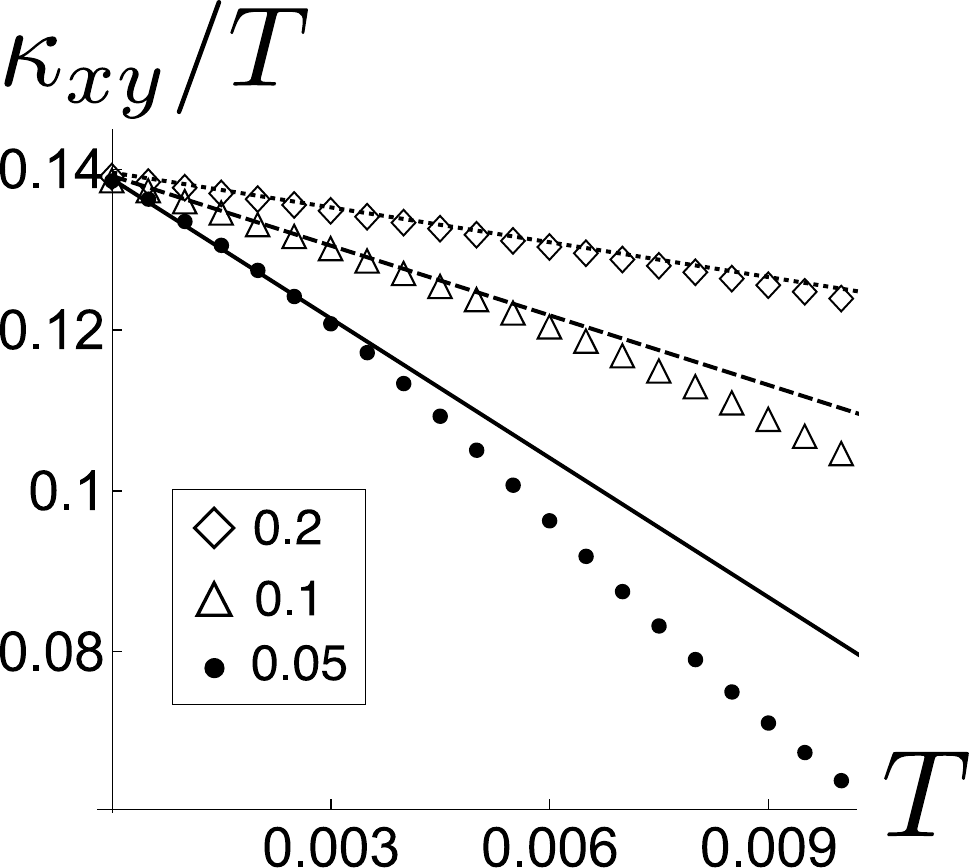}}\\
      (b) 
   \end{center}
  \end{minipage}
 \end{tabular}
    \caption{\label{hexad_A_result}   (a)(b) Dependence of $\La$ on $\e$ and $\kxy/T$ on $T$for \dxxyy-wave SC with A-type FS for model of Eq.(\ref{tb-d6h}) with the parameters $t_2/t_1 = 1.5, \mu/t_1 = -8$. 
    {The unit for $\epsilon$ and $T$ is identical as declared before. }
    The filled circle, open triangle, and open diamond are numerical results for $\Delta/k_F^n = 0.05, 0.1, 0.2$, respectively, for the gap function of Eq. (\ref{hexag_gap-d}). The solid, dashed, and dotted lines correspond to the fit using Eq. (\ref{kxy-pn2}). }

 \end{center}
\end{figure}
\begin{figure}[h]
 \begin{center}
 \begin{tabular}{c}

  \begin{minipage}{0.45\hsize}
    \begin{center}
     \resizebox{!}{0.9\hsize}{\includegraphics{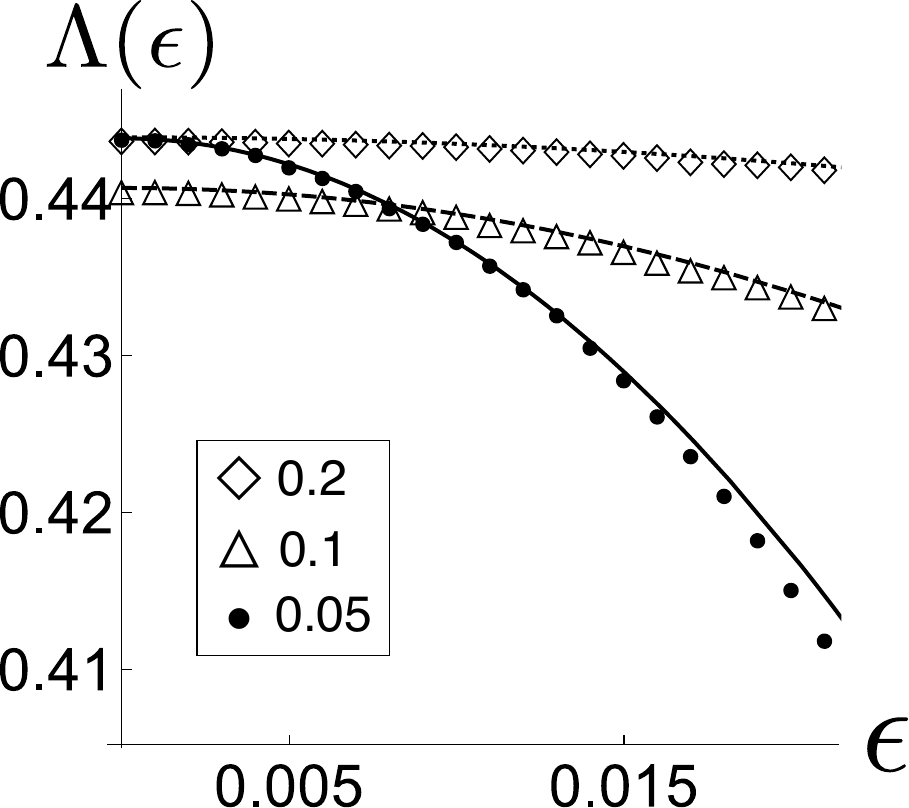}}\\
 (a) 
   \end{center}
  \end{minipage}
\hskip 0.5 cm
  \begin{minipage}{0.45\hsize}
   \begin{center}
     \resizebox{!}{0.9\hsize}{\includegraphics{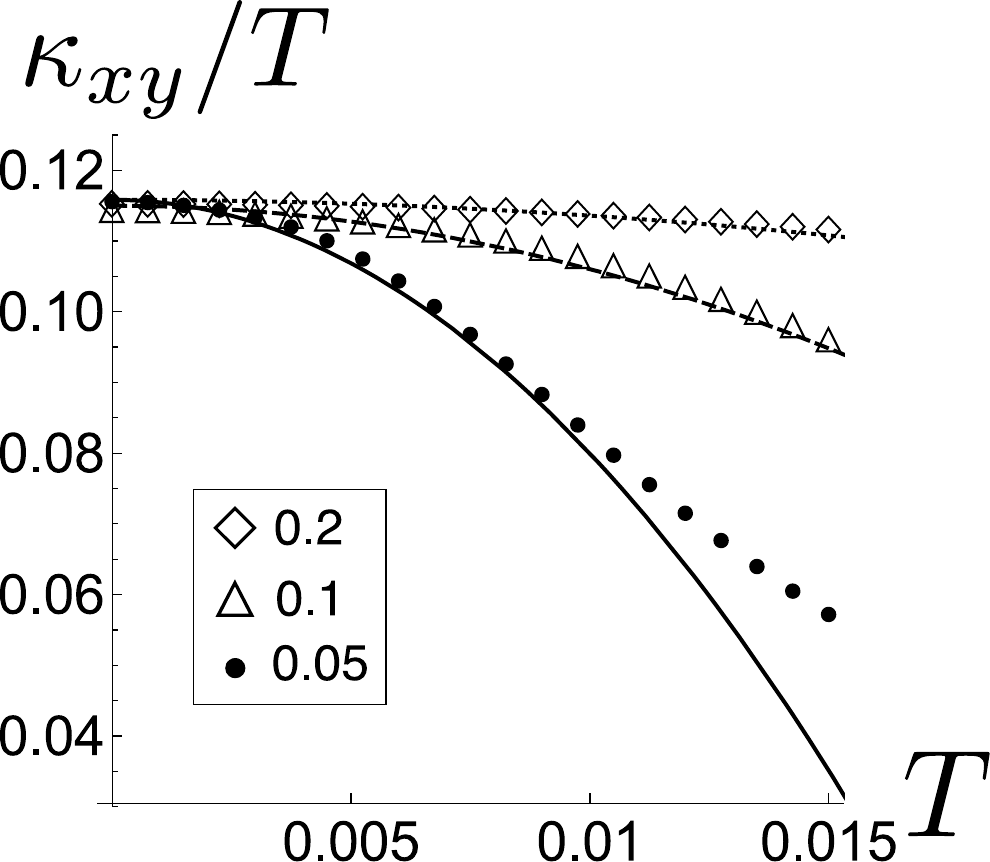}}\\
 (b) 
    \end{center}
  \end{minipage}

 \end{tabular}
    \caption{\label{hexad_C_result}  (a)(b) Dependence of $\La$ on $\e$ and $\kxy/T$ on $T$for \dxxyy-wave SC with C-type FS for model of Eq. (\ref{tb-d6h}) with the parameters $t_2/t_1 = -1.5, \mu/t_1 = 5.3$. 
    {The unit for $\epsilon$ and $T$ is identical as declared before. }
    The filled circle, open triangle, and open diamond are numerical results for $\Delta/k_F^n = 0.05, 0.1, 0.2$, respectively, for the gap function of Eq. (\ref{hexag_gap-d}). The solid, dashed, and dotted lines correspond to the fit using Eq. (\ref{kxy-pn1}).}
 \end{center}
\end{figure}

Different winding number $ n $ of the point node is reflected in the different power laws. We can compare this feature again with numerical calculation for specific models on a hexagonal lattice with a \dxxyy-wave superconducting state. For the purpose of illustration, we choose the FS of type A and C in Fig. \ref{hexa_fermi} (a) and (c), respectively.  For the type A FS the point nodes are expected to have winding $ n= \pm 2 $ along the line $ k_x = k_y=0 $. On the other hand, for the type C FS, point nodes at the BZ boundary have winding number $ n = \pm 1 $. Thus, we expect different power law behavior in $ \La(\epsilon) $ according to our discussion for these two cases. 

In Fig. \ref{hexad_A_result} and Fig. \ref{hexad_C_result}, we display the numerical result of $\La(\e)$ and $\kxy/T$ for a type A and type C FS, respectively. As is expected, the fitting lines correspond with  the results very well, strongly supporting our analytical description. Here, we determine $k_F$ from the kinetic energy, Eq. (\ref{tb-d6h}), as in the previous section, and $\Delta_0$ from the expansion coefficient of Eq. (\ref{hexag_gap-d}).

\section{Discussion}

The spontaneous thermal Hall effect in chiral superconductors conveys specific information. First of all, the observation of a non-vanishing value of $\kxy^L$ can be considered as evidence for chirality of a superconductor and the geometry of the measurement allows us to determine the chiral axis. 
We have seen that the $\kxy/T$ in the limit $ T \to 0 $ ($\kxy^L/T$) provides input on the the Chern number combined with the projected distance of Weyl points connected by Fermi arcs on the surface. Combined with other measurements of the Fermi arcs, e.g. ARPES, there is even a chance to determine $k_z$-dependent Chern number $ C(k_z) $ to further characterize the chiral superconducting phase. The deviation from the limiting value at $ T=0 $, $\Delta\kxy$, contributes insight into the nodal structure of the gap through the temperature dependence, which has the form of power laws for nodal gaps.
Note that such power laws are also known from other, more readily accessible, quantities such as the specific heat, London penetration depth, NMR $1/T_1 $ and ultrasound absorption rate. 
Naturally the analysis of experimental results would be rather involved in case of a multi-band superconductor, as is usually the case for real materials.

A comment to the quantitative level is in order here. The magnitude of $ \kxy^L/T$ is estimated by
\begin{equation}
\frac{\kxy^L}{T} \sim \frac{\pi}{12} \La(0) \frac{k_B^2}{\hbar} \sim 10^{-4} - 10^{-3} \frac{\rm W}{\rm K^2 m}
\end{equation}
for $ \La(0) \sim k_F/\pi \sim 1/d $ with a $c$-axis lattice constant of $ d \sim 1 nm $ (layer distance). This is not easy to resolve at very low temperatures as necessary for the candidate chiral superconductors listed below. We then consider the relative magnitude of the correction $\Delta\kxy / T $ for the case
of a nodal line in the gap as given by Eq. (\ref{kxy_LN}),
\begin{equation}
\left| \frac{\Delta\kxy}{\kxy^L} \right| \approx \frac{27 \zeta(3)}{\pi^2} \frac{M_{\rm LN}}{\La(0)} T \sim \frac{T}{\Delta_0} ,
\end{equation}
and for point nodes of charge $n=1$ following Eq. (\ref{kxy-pn1}),
\begin{equation}
\left| \frac{\Delta\kxy}{\kxy^L} \right| \approx \frac{84 \pi^2 }{30} \frac{M_{\rm PN}^{(1)}}{\La(0)} T^2  \sim \frac{T^2}{\Delta_0^2} .
\end{equation}
This result is per nodal point or line and estimates the contamination of the zero-temperature limit by quasi-particle contamination through thermal activation. 
Note that, in case of a nodeless gap, this ratio is given by
\begin{equation}
\left| \frac{\Delta\kxy}{\kxy^L}  \right| \sim e^{-\Delta_0/T} .
\end{equation}

Let us briefly address a few examples of potential chiral superconductors. Sr$_2$RuO$_4$ and URu$_2$Si$_2$ share the tetragonal symmetry of their crystal structure. The first is considered to be a chiral $p$-wave superconductor of the type \pxy \cite{luke_1998,xia_2006,mackenzie_2003}, and the second a chiral $ d $-wave superconductor with \dzxzy - pairing structure\cite{kasahara_2007,goswami_2013}. The gap structures impose point nodes on the first and point and line nodes on the second, in the generic case with A-type of Fermi surfaces. Note that Sr$_2$RuO$_4$ has a quasi-two-dimensional band structure with three bands being of B-type\cite{mackenzie_2003}. Thus, in this latter case, we would not expect power law behavior for temperature dependence of $\Delta\kxy$, but rather an exponential type, whereby the gap may actually be rather small as the minimal gap value is relevant. All these states would in their simplest realization give rise to $ C(k_z) = 0 , \pm 1 $ on a given Fermi surface. 

Hexagonal symmetry is the underlying crystal structure for the candidates UPt$_3$ and SrPtAs. The heavy Fermion superconductor UPt$_3$ has a double transition breaking time reversal symmetry below the second transition. For the low-temperature phase various chiral pairing states have been proposed: 
the chiral $p$-wave state \pxy, the two chiral $f$-wave state, $f_{x(5z^2-r^2)}+if_{y(5z^2-r^2)}$ and $f_{z(x^2-y^2)}+if_{zxy}$ belonging to the representations $E_{1u} $ and $ E_{2u} $ respectively\cite{izawa_2014}. Note that the $ E_{1u} $ state (including also the chiral $p$-wave state) has minimal  $ C(k_z) = 0 , \pm 1 $ on a single Fermi surface. On the other hand, the $E_{2u} $ yields like the \dxxyy -state Chern numbers up to 2. Also SrPtAs, potentially a chiral $d$-wave superconductor of the type \dxxyy within the representation $E_{2u} $ \cite{biswas_2013,fisher_2014}, belongs to this class. 

Eventually we would also like to mention that the model cases discussed in section \ref{MCS} represent the most simple versions of chiral superconductors. We modify here the case of the chiral \pxy-wave superconductor by changing from in-plane nearest-neighbor pairing in Eq.(\ref{tetrad_gap-p}) to next-nearest neighbor pairing where $ \vec{R}_{ij} = (\pm a,\pm a,0) $ leading to
\begin{equation}
\Delta_{\vk} = \Delta_0 \left( \cos k_y \sin k_x + i \cos k_x \sin k_y  \right) .
\end{equation}
This state has new zero-lines in the Brillouin zone, 
\begin{equation} \begin{array}{ll}
\vk_{0,1} = (0,0,k_z), (\pi, \pi, k_z),  & n=1, \\
\vk_{2,3} = (\pi,0,k_z), (0, \pi,k_z), & n=1, \\
\vk_{4-7} = (\pm \pi/2,\pm \pi/2,k_z ), &  n = -1.
\end{array}
\end{equation}
This new lines of zero can lead to additional structure in $ C(k_z)$, if they are encircled by the Fermi surface in BZ($k_z$). For a A-type FS there can be a range of 
$k_z $ for which $ C(k_z) = - 3 $,
\begin{equation}
C(k_z) = \begin{cases} -3 & | k_z | < k' \\ +1 & k' < |k_z| < k'' \\ 0 & k'' < | k_z | \leq \pi \end{cases}
\end{equation}
which generates, by bulk-edge correspondence, four new Fermi arcs of negative orientation that connect the new Weyl points at $ k_z = \pm k' $. This structure 
naturally complicates the analysis of $ \kxy $, as not only $ \La(0) $ changes but also $ \Delta \kxy $ involves eight additional point nodes. The observation of the Fermi arcs by means of ARPES, however, could help to anticipate the complication.

\section{Conclusions}

The spontaneous thermal Hall effect in (quasi) two-dimensional chiral superconductors is expected to be universally quantized at low temperatures, reflecting the topology of this phase\cite{read_2000,sumiyoshi_2013}. In three-dimensional superconductors this is not the case anymore, because a chiral superconductor is not anymore a topological phase due to zero nodes in the excitation gap, imposed by symmetry and gap structure. While the universal quantization is not realized anymore, we nevertheless expect a spontaneous thermal Hall effect. In this paper we have analyzed what information is contained in the signal $ \kxy $ as a function of temperature. 

The value of $ \kxy / T  $ in the zero-temperature limit ($ \kxy^L/T$) provides information about the topological structure of the pairing state. The value of $ \kxy^L/T$ consists of the contributions of all Fermi arcs of the edge states connecting projected Weyl point at a given surface. Each point on a Fermi arc contributes $ \pm \pi /12 $ to $ \kxy^L/T$ which, together with the knowledge of the Fermi arc structure by other experiments, can be used to extract information about the structure of the gap function. The nodal structure is reflected by the deviation of $ \kxy / T  $ from $ \kxy^L/T$ as a function of temperature. As this is due to the quasiparticle excitations the nodal states govern the behavior and lead to a power law dependence in $T$. The power is determined by the structure of the nodes, whether they are line or point nodes and by the winding number in the case of point nodes.

We intentionally avoid the discussion of the effect of external magnetic fields here as this involves various aspects which are beyond the scope of our paper \cite{simon_1997,kubert_1998}. Magnetic fields can, however, be an issue because of the requirement that the superconducting phase is single domain. The fact that both chiralities are energetically degenerate in zero magnetic field, introduces the possibility of chiral domain formation which would spoil our discussion. On the other hand, undergoing the superconducting phase transition in a magnetic field parallel to the chiral axis would be one way to achieve the single domain situation. For a clean experimental investigation it would have to be carefully checked that the measurements would not be contaminated by trapped flux. 

Despite all complications, we believe that the measurement of the thermal Hall effect would be a neat consistency check for the existence of chiral Cooper pairing in time reversal symmetry breaking superconducting phases. The hope is that in future chiral superconductors would be discovered with a considerably higher critical temperature than seen in the examples mentioned above.

\section*{Acknowledgements}
The authors wish to thank H. Katsura, T. Sakakibara, and T. Shibauchi for many fruitful discussions. This work was financially supported by a grant of the Swiss National Science Foundation. N.Y. was supported by the JSPS through Program for Leading Graduate Schools (ALPS) and JSPS fellowship (JSPS KAKENHI Grant No. JP17J00743). 
N.Y. is grateful for the hospitality of the Institute for Theoretical Physics and the Pauli Centre of ETH Zurich. 

\appendix
\section{single-band Berry curvature}\label{berrycal_app}
Here we calculate the Berry curvature in a two-level system \cite{xiao_2010} in order to justify the expression in Eq. (\ref{berry_single}). The generic Hamiltonian dependent on an external parameter $\vk = (k_x, k_y,\dots)\in \mathbb{R}^n$ ($n\geq2$) takes the form,
\begin{eqnarray}
H=\vec{h}(\vk)\cdot\vec{\sigma},
\end{eqnarray}
where the $\vec{\sigma}=(\sigma_x,\sigma_y, \sigma_z)$ is the Pauli matrix. 
Parameterizing the vector as $\vec{h}(\vk) = (\sin\theta_{\vk}\cos\phi_{\vk},\sin\theta_{\vk}\sin\phi_{\vk},\cos\theta_{\vk})$, the eigenstates with energy $\pm|\vec{h}(\vk)|$ can be concisely expressed as
\begin{eqnarray}
\ket{u_-(\vk)} = \left(
\begin{array}{c}
\sin\frac{\theta_{\vk}}{2}e^{-i\phi_{\vk}}\\
\cos\frac{\theta_{\vk}}{2}
\end{array}
\right),\\
\ket{u_+(\vk)} = \left(
\begin{array}{c}
\cos\frac{\theta_{\vk}}{2}e^{-i\phi_{\vk}}\\
\sin\frac{\theta_{\vk}}{2}
\end{array}
\right).
\end{eqnarray}
For simplicity, we drop the index $\vk$ in the following. 
The Berry connection of the lower (occupied) band is given by
\begin{eqnarray}
\mathcal{A}_{\theta}^- &=&\braket{u_-|i\partial_{\theta}u_-} = 0,\\
\mathcal{A}_{\phi}^- &=&\braket{u_-|i\partial_{\phi}u_-} = \sin^2\frac{\theta}{2},\\
\end{eqnarray}
and accordingly the Berry curvature is 
\begin{eqnarray}
\Omega_{\theta\phi}^- = \partial_{\theta}\mathcal{A}_{\phi}^- - \partial_{\phi}\mathcal{A}_{\theta}^- = \frac{1}{2}\sin\theta.
\end{eqnarray}
By simply considering the variable transformation, the Berry curvature in the Cartesian coordinate is given by
\begin{eqnarray}
\Omega_{k_x k_y}^- = \frac{\partial(\theta,\phi)}{\partial (k_x,k_y)}\cdot \Omega_{\theta\phi}^- = \frac{1}{2}\frac{\partial(\phi, \cos\theta)}{\partial(k_x,k_y)}.
\end{eqnarray}
The elements of the Jacobian can be expressed as
\begin{eqnarray}
\frac{\partial\phi}{\partial k_i} &=& \frac{h_1\partial_i h_2-h_2\partial_i h_1}{|\vec{h}|^2-h_3^2},\\
\frac{\partial \cos\theta}{\partial k_i} &=& \frac{(|\vec{h}|^2-h_3^2)\partial_i h_3 - h_3 (h_1\partial_ih_1 + h_2 \partial_i h_2)}{|\vec{h}|^3},
\end{eqnarray}
where $\partial_i$ is the partial derivative with respect to $k_i$, and thus the Berry curvature is 
\begin{eqnarray}
\Omega_{k_x k_y}^- &=& \frac{1}{2}\kl \frac{\partial\phi}{\partial k_x}\frac{\partial(\cos\theta)}{\partial k_y} -\frac{\partial\phi}{\partial k_y}\frac{\partial(\cos\theta)}{\partial k_x} \kr\\
 &=& \frac{1}{2}\frac{\vec{h}\cdot (\partial_{k_x} \vec{h}\times \partial_{k_y} \vec{h})}{|\vec{h}|^3}.
\end{eqnarray}
In particular, the Berry curvature for single-band BdG Hamiltonian in $k$-space is obtained by substituting $\vec{h} = (m_1, -m_2, m_3)$ with $\vec{m}\ksp =(\mathrm{Im}\Delta\ksp, \mathrm{Re}\Delta\ksp, \e\ksp)$ as
\begin{eqnarray}
\Omega_{k_xk_y}^{-} = \frac{1}{2}\frac{\vec{m}\ksp\cdot (\partial_{k_x} \vec{m}\ksp\times \partial_{k_y} \vec{m}\ksp)}{|\vec{m}\ksp|^3}.
\end{eqnarray}

\bibliographystyle{apsrev4-1}
\bibliography{TSCTHE_bibtex}
\end{document}